\documentclass[fleqn,usenatbib]{mnras}

\usepackage[T1]{fontenc}
\usepackage{ae,aecompl}

\usepackage{newtxtext,newtxmath}
\usepackage{graphicx}
\usepackage{color}
\usepackage{soul}
\usepackage{url}
\usepackage{bm}         
\usepackage{times}
\usepackage{dcolumn}
\usepackage{bm}
\usepackage{epsf}
\usepackage{amssymb}
\usepackage{caption}
\usepackage{subcaption}
\newcommand{\be}{\begin{equation}}
\newcommand{\ee}{\end{equation}}

\newcommand{\beq}{\begin{equation}}
\newcommand{\eeq}{\end{equation}}
\newcommand{\beqn}{\begin{eqnarray}}
\newcommand{\eeqn}{\end{eqnarray}}

\usepackage{color}

\begin{document}

\label{firstpage}

\title[FRBs from decelerating blast waves]{Fast radio bursts as synchrotron maser emission from decelerating relativistic blast waves}

\author[]{Brian D.~Metzger, Ben Margalit$^{2,4}$, Lorenzo Sironi$^{3}$\\
$^{1}$Department of Physics, Columbia University, New York, NY, 10027, USA\\
$^{2}$Department of Astronomy and Theoretical Astrophysics Center, University of California, Berkeley, CA, 94720, USA\\
$^{3}$Department of Astronomy, Columbia University, New York, NY, 10027, USA\\
$^{4}$NASA Einstein Fellow\\
}

\maketitle

\begin{abstract}
Fast radio bursts (FRB) can arise from synchrotron maser emission at ultra-relativistic magnetized shocks, such as produced by flare ejecta from young magnetars.  We combine particle-in-cell (PIC) simulation results for the maser emission with the dynamics of self-similar shock deceleration, as commonly applied to gamma-ray bursts (GRB), to explore the implications for FRBs.  The upstream environment is a mildly relativistic baryon-loaded shell released following a previous flare, motivated by the high electron-ion injection rate $\dot{M} \sim 10^{19}-10^{21}$ g s$^{-1}$ needed to power the persistent radio nebula coincident with the repeating burster FRB 121102 and its high rotation measure.  The radio fluence peaks once the optical depth ahead of the shock to induced Compton scattering $\tau_{\rm c} \lesssim 3$.  Given intervals between major ion ejection events $\Delta T \sim 10^{5}$ s similar to the occurrence rate of the most powerful bursts from FRB 121102, we demonstrate the production of $\sim 0.1-10$ GHz FRBs with isotropic radiated energies $\sim 10^{37}-10^{40}$ erg and durations $\sim 0.1-10$ ms for flare energies $E \sim 10^{43}-10^{45}$ erg.   Deceleration of the blast wave, and increasing transparency of the upstream medium, generates temporal decay of the peak frequency, similar to the observed downward frequency drift seen in FRB 121102 and FRB 180814.J0422+73.  The delay $\Delta T \gtrsim 10^{5}$ s between major ion-injection events needed to clear sufficiently low densities around the engine for FRB emission could explain prolonged "dark periods" and clustered burst arrival times.  Thermal electrons heated at the shock generate a short-lived $\lesssim 1$ ms (1 s) synchrotron transient at gamma-ray (X-ray) energies, analogous to a scaled-down GRB afterglow. 
\end{abstract}
\begin{keywords}
Shock waves -- stars:neutron  -- radio continuum:transients
\end{keywords}

\section{Introduction}

Fast radio bursts (FRB) are luminous pulses of $\sim$ GHz radio emission with durations of less than a few milliseconds and large dispersion measures (DM), indicating an extragalactic origin (\citealt{Lorimer+07,Keane+12,Thornton+13,Spitler+14,Ravi+15,Champion+16,Petroff+16,Lawrence+17,Shannon+18,James+18}). The cosmological origin of at least one burster was confirmed by the discovery of the repeating source FRB 121102 \citep{Spitler+14,Spitler+16,Scholz+16,Law+17} and its localization \citep{Chatterjee+17} to a dwarf star-forming galaxy at redshift $z = 0.1927$ \citep{Tendulkar+17}.  A second repeating source was recently discovered by the CHIME survey \citep{CHIME+19b}.  A larger sample of FRBs, including new repeating sources, will be discovered over the next few years by surveys including SUPERB \citep{Keane+18}, CHIME \citep{CHIME+18}, and ASKAP \citep{Shannon+18}.  

The short durations of FRBs, with sub-structure down to tens of microseconds \citep{Michilli+18}, are suggestive of their central engines being stellar-mass compact objects such as pulsars \citep{Cordes&Wasserman16} or magnetars (e.g.~\citealt{Popov&Postnov13,Lyubarsky14,Kulkarni+15}), especially ones at particularly active stages in their lives \citep{Metzger+17,Beloborodov17}.  However, other engine scenarios remain in contention which can in principle produce repeating bursts, such as the collision between primordial magnetic dipoles \citep{Thompson17} or "cosmic combs" produced by the interaction between a neutron star's magnetosphere and a dense outflow from a nearby AGN \citep{Zhang17,Zhang18}.  

The high fluxes $\sim 0.1-1$ Jy and large distances of FRBs imply enormous brightness temperatures $\gtrsim 10^{37}$ K, requiring a coherent emission process (e.g.~\citealt{Katz16,Lyutikov19}).  The two most commonly discussed mechanisms are curvature radiation produced close to the surface of the neutron star (e.g.~\citealt{Kumar+17,Lu&Kumar18}) and the maser synchrotron process (e.g.~\citealt{Hoshino&Arons91,Long&Peer18}).  A common variant of the latter postulates emission from an ultra-relativistic shock moving towards the observer, which propagates into an upstream medium of moderately high magnetization, $\sigma \gtrsim 10^{-3}$ \citep{Lyubarsky14,Beloborodov17}.\footnote{The shock magnetization $\sigma$ is defined as the ratio of incoming Poynting flux to particle energy flux.}  Such shocks are mediated by Larmor rotation of charges entering the shock and gyrating around the ordered magnetic field.  This creates the necessary population inversion in the form of an unstable ring-like particle distribution function, which relaxes by transferring energy into an outwardly propagating coherent electromagnetic wave (e.g.~\citealt{Gallant+92,hoshino_92,Amato&Arons06,Hoshino08,Sironi&Spitkovsky09,Sironi&Spitkovsky10,Iwamoto+17,Iwamoto+18,Plotnikov&Sironi19}).  In the magnetar scenario, these shocks result from the transient release of energy during the earliest stages of a flare.  Part of the star's magnetosphere ``snaps off'' while still relatively clean of plasma, transforming into an outgoing $\sigma \gg 1$ magnetic pulse that collides with the surrounding environment on much larger radial scales \citep{Lyubarsky14}.

The magnetar-powered synchrotron maser shock model makes several predictions which are consistent with FRB observations.  First, it explains the high measured linear polarization of some FRBs \citep{Ravi+16,Petroff+17,Caleb+18}, which for FRB 121102 is nearly 100\% \citep{Michilli+18,Gajjar+18}.\footnote{Although some FRBs show no detectable linear polarization, this may be the result of propagation effects in a local magnetized medium, such as Faraday rotation (which cannot be subtracted off without sufficient spectral resolution; \citealt{Michilli+18}) or Faraday conversion into circularly polarized emission \citep{Vedantham&Ravi18,Gruzinov&Levin19}.}  Particle-in-cell (PIC) simulations show that a large-amplitude linearly polarized X-mode wave (the nascent FRB) is created at the shock front and propagates into the upstream medium (e.g.~\citealt{Gallant+92,Plotnikov&Sironi19}).  The intrinsic polarization angle of bursts from FRB 121102 was measured to be roughly constant over $\gtrsim 7$ months of observations \citep{Michilli+18}, during which the source presumably underwent thousands of bursts.  This requires a fixed direction for the magnetic field of the upstream plasma into which the FRB-producing ejecta collides.  Such a fixed field structure naturally occurs in the outflow from a rotating compact object, for which the magnetic field wraps around the (approximately fixed) rotation axis. 

Another appealing aspect of the synchrotron maser is its high efficiency, $f_{\xi}$, for converting the kinetic energy of the ejecta into coherent electromagnetic radiation.  One-dimensional PIC simulations of magnetized ultra-relativistic shocks propagating in {\it pair plasmas} find a maximum efficiency of up to several percent for an upstream magnetization $\sigma \sim 0.1$, which decreases as $f_{\xi} \propto \sigma^{-2}$ for $\sigma \gg 1$ \citep{Plotnikov&Sironi19}.  This is compatible with the lower limit on the efficiency of $f_{\xi} \gtrsim 10^{-6}-10^{-7}$ for FRB 121102 in magnetar models, under the assumption that the source bursts in a (time-averaged) isotropic manner with its current luminosity function for an active lifetime of $\sim 100$ yr \citep{Nicholl+17}.  Although the maser efficiency could be lower in the physical case of higher dimensions  than was found in 1D PIC simulations \citep{Sironi&Spitkovsky09},  recent multi-dimensional simulations find the drop in efficiency is only a factor of $\lesssim 10$ for $\sigma\lesssim 1$ (\citealt{Iwamoto+17,Iwamoto+18}) and even less at high magnetizations $\sigma\gtrsim 1$ (Sironi et al, in prep.). The properties of the synchrotron maser  in {\it electron-ion} and {\it pair-ion} plasmas are less well characterized: while one-dimensional simulations find efficient electron (and positron) maser emission \citep{Hoshino08}, in multi-dimensional studies it has been shown that efficient maser emission leads to strong heating of the incoming pairs \citep{lyubarsky_06}, which in turns suppresses the efficiency of the synchrotron maser \citep{Sironi&Spitkovsky10}.

For low magnetizations ($\sigma \lesssim 1$), the spectral energy distribution (SED) of the synchrotron maser is peaked in the post-shock frame at a few times the plasma frequency of the upstream electrons $\nu_{\rm pk} \sim 3\nu_{\rm p}$, with power extending to frequencies $\gg \nu_{\rm p}$ and detailed structures due to overlapping line-like features produced by a large number of resonances \citep{Plotnikov&Sironi19}.\footnote{In electron-ion or pair-ion plasmas, the synchrotron maser emission propagating upstream has the effect of boosting the incoming electrons (and positrons) towards the shock \citep{lyubarsky_06}, so they enter the shock with bulk kinetic energy comparable to the incoming ions, which leads to maser emission peaking near the ion plasma frequency.}  These features are at least qualitatively consistent with the observed complex, and sometimes narrow-band SEDs, of observed FRBs (e.g.~\citealt{Ravi+16,Law+17,Macquart+18}).  Temporal-frequency evolution could be imprinted by plasma lensing effects during the propagation to Earth (e.g.~\citealt{Cordes+17,Main+18}) rather than being an intrinsic property of the bursts.  Furthermore, induced scattering by the matter just upstream of the shock \citep{Lyubarsky08} could play a crucial role in shaping the observed light curve and spectrum.

A key question in the shock-powered FRB scenario is the nature of the upstream medium into which the ultra-relativistic ejecta from the engine collides.  Relevant here is the compact unresolved ($< 0.7$ pc) luminous persistent synchrotron radio source located coincident with the spatial position of FRB 121102 \citep{Chatterjee+17,Marcote+17}.  A related clue is the enormous rotation measure of the bursts, RM $\sim 10^{5}$ rad m$^{-2}$ \citep{Michilli+18}.  The persistent emission and high-RM likely originate from the same medium, showing that the FRB source is embedded in a dense magnetized plasma (e.g.~\citealt{Michilli+18,Vedantham&Ravi18}).  While this environment need not be directly related to the bursting source itself (for instance if a flaring magnetar just happens to reside close to an AGN; \citealt{Eatough+13}), it could instead be a compact transient nebula powered by the FRB central engine \citep{Murase+16,Metzger+17,Beloborodov17,Waxman17}.  The high RM would then indicate that the ejecta from the bursting source is predominantly of an ion-electron composition by particle number as well as mass (an electron/positron pair plasma, such as those of normal rotational-powered pulsar winds, contributes no net RM).

\citet{Margalit&Metzger18} demonstrate that a single expanding and continuously-energized magnetized ion-electron nebula embedded within a young supernova remnant of age $10-40$ years is consistent with all of the properties of the persistent source of FRB 121102 (size, flux, self-absorption constraints) and the large but decreasing RM (see also \citealt{Margalit+18}).  The persistent emission can be explained as synchrotron radiation from electrons heated thermally at the termination shock (of size $\sim 10^{17}$ cm) of the magnetar wind behind the expanding supernova ejecta, while the RM originates from the electrons injected earlier in the nebula's history and cooled through expansion and radiative losses to become non-relativistic.  Of particular relevance to this work, the properties of the ion-electron injection are relatively tightly constrained: the time-averaged wind entering the nebula must possess a sub-relativistic velocity $v_w \sim 0.5$ c (similar to the escape speed of a neutron star) and a present-day mass injection rate of $\dot{M} \sim  10^{19}-10^{21}$ g s$^{-1}$.

The high required time-averaged baryon loading of material entering the nebula contrasts with the much "cleaner" but short-lived $\lesssim 1$ ms ultra-relativistic ejection events needed to power FRBs themselves.  This suggests a picture in which the bulk of the ions emerge from the star after major flares, and then subsequently serves as the upstream medium into which the next flare collides to produce the FRB, as first proposed by \citet{Beloborodov17}.  One is thus led to hypothesize that FRBs, at least those from FRB 121102, result from internal shocks in the magnetar wind between two media with rather different properties.

This paper develops the internal-shock scenario for FRB in light of recent particle-in-cell (PIC) simulation work on the properties of the synchrotron maser emission \citep{Plotnikov&Sironi19}.  Motivated by observations of the well-studied repeating source FRB 121102, we then apply our results to address several outstanding questions, including:
\begin{itemize}
\item{What determines the $\sim 0.1-10$ GHz frequency range over which FRBs are detected?  Does this range arise naturally from the model, or is fine-tuning of the upstream medium required?}
\item{If FRBs originate from sudden reconnection events in neutron star magnetospheres, for which the light crossing time is $\lesssim 0.1$ ms, then how can bursts possess intrinsic durations up to several milliseconds ($\gtrsim 10$ times longer than would naively be guessed)?}
\item{Spectral features observed during sub-bursts from FRB 121102 drift downwards in frequency over time \citep{Hessels+18}.  Similar behavior was seen from the new CHIME repeating source, FRB 180814.J0422+73 \citep{CHIME+19b}.  What produces this drifting?  Why is it always downward? and how does it inform the nature of the shocks and upstream environment? }
\item{Time-resolved observations of the bursts from FRB 121102 show a narrowly-peaked spectral energy distribution of width $\Delta \nu/\nu \sim 0.1-0.2$ \citep{Law+17}.  If FRB emission arises from a relativistic shock then an intrinsically wider spectrum would imprinted by Doppler smearing across the shock front.  If FRBs arise from relativistic shocks, attenuation or amplification of the intrinsic spectrum by an external medium is likely playing an important role.  What is this external medium and its effect on the observed radiation?}
\item{Given the efficiency of FRB emission, accounting for attenuation by the upstream medium, what energetic constraints are imposed on the central engine by the repeater FRB 121102?}
\item{The burst arrival times from FRB 121102 are non-Poissonian and clustered (e.g.~\citealt{Spitler+14,Opperman+18,Katz18,Li+19}), with large "dark" phases of little or no apparent FRB activity (e.g.~\citealt{Price+18}).  Does this behavior indicate true intermittency of the central engine activity, or can it result also from a time-changing external environment (e.g. as shaped by prior flares)?}
\item{FRB 121102 showed an increase of $\sim 1-3$ pc cm$^{-3}$ in its DM over a 4 year baseline \citep{Hessels+18}.  While an expanding ionized supernova ejecta shell can result in a time-dependent DM \citep{Connor+16,Piro16}, the predicted evolution is usually a {\it decrease} \citep{Margalit+18} unless the medium surrounding the supernova is unusually dense \citep{Yang&Zhang17,Piro&Gaensler18}.  What additional mechanisms can give rise to stochastic or secular variation in DM?}
\item{Any coherent maser synchrotron emission should have accompanying {\it incoherent} synchrotron radiation at much higher photon energies from electrons thermally heated at the same shock \citep{Lyubarsky14}.  What are the properties of this multi-wavelength FRB afterglow?}
\end{itemize}

Although the magnetar scenario is appealing for several reasons, a wider range of models postulate a central engine that impulsively injects energy into a dense external environment (e.g. the gaseous environment of an AGN).  This is particularly relevant given that it is not clear whether the nature and environment of FRB 121102 is generic to all repeaters or to the broader class of FRBs which have thus far been observed to burst only once \citep{Caleb+18}.  This motivates developing the more general scenario of a decelerating ultra-relativistic blast wave and its time-dependent synchrotron maser emission.  This interaction was first pioneered in the context of AGN jets (e.g.~\citealt{Blandford&McKee76}) and the synchrotron afterglow of gamma-ray burst (GRB) jets (e.g.~\citealt{Meszaros&Rees93,Katz94,Sari&Piran95,Sari+98}).  Here we extend this analysis to FRBs, providing scaling relationships that should prove useful in modeling future events in terms independent of the central engine model.

\begin{figure*}
\centering
\includegraphics[width=1.0\textwidth]{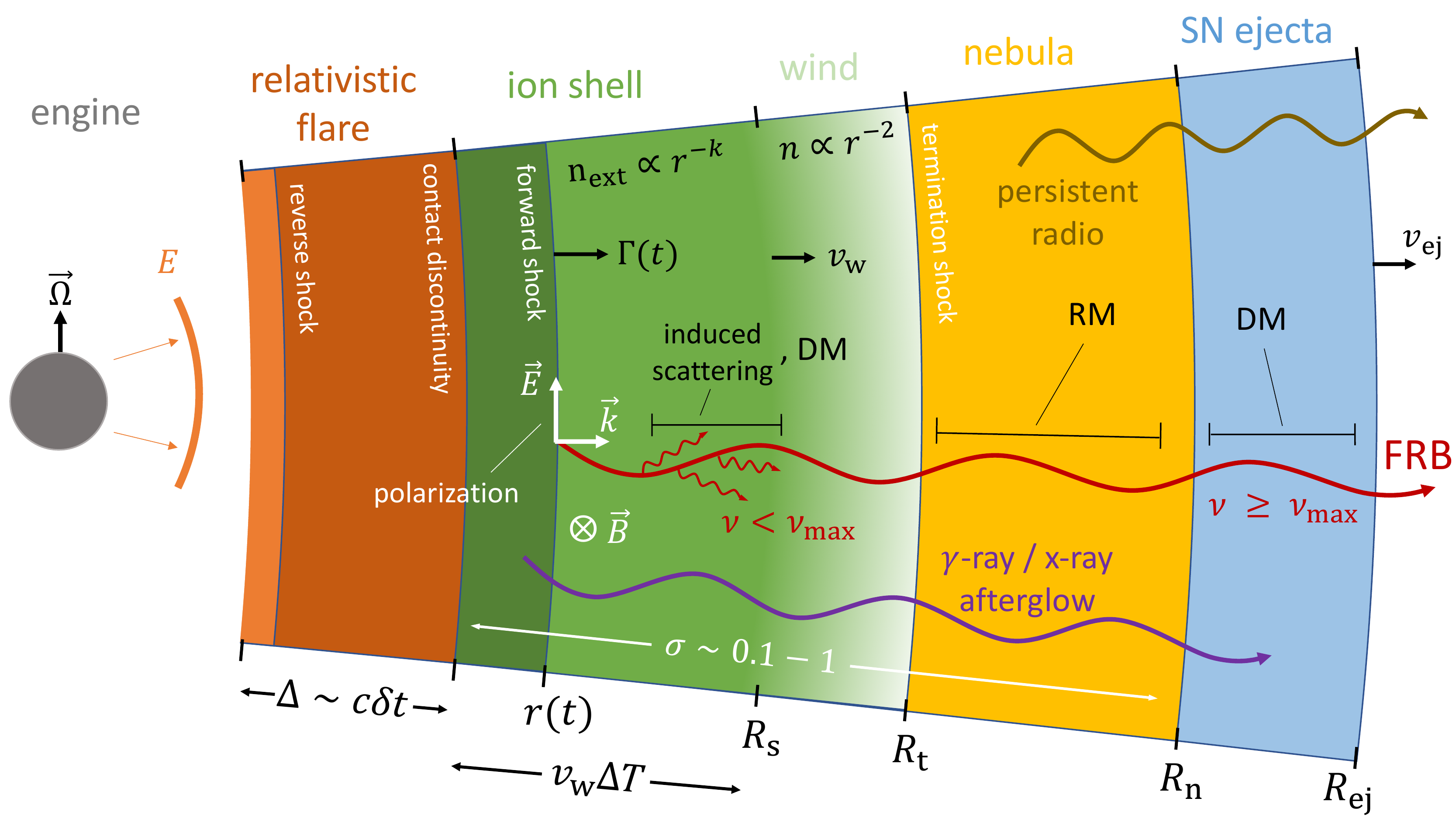}\\
\caption{Radial scales and physical processes surrounding a repeating FRB source as described in this paper.  The central engine releases an ultra-relativistic shell of energy $E$, duration $\delta t \lesssim 1$ ms, and radial width $c \delta t$, which collides with a mildly-relativistic magnetized ion-electron shell of velocity $v_w$, baryon density $n_{\rm ext} \propto r^{-k}$, magnetization $\sigma \sim 0.1-1$ and total width $v_w \Delta T$, as released following the previous major flare a time $\Delta T$ ago.  The shell  decelerates through reverse and forward shocks ($\S\ref{sec:dynamics}$), the latter of which produces the observed coherent radio emission (fast radio burst) through the synchrotron maser mechanism ($\S\ref{sec:FRB}$; Fig.~\ref{fig:SED}).  The upstream magnetic field $\vec{B}$ is wrapped in the toroidal direction perpendicular to the rotation axis $\Omega$ of the central engine, resulting in linear polarization of the FRB emission along the direction of $\Omega$.  The radio pulse is attenuated in the ion shell by induced Compton scattering at low frequencies $\nu < \nu_{\rm max}$ ($\S\ref{sec:scattering}$; eq.~\ref{eq:numax}).  As the blast wave decelerates, the decreasing Lorentz factor $\Gamma$ of the shocked gas and the reduced scattering optical depth of the upstream medium results in a downward drift of $\nu_{\rm max}$ over the duration of the observed burst (Fig.~\ref{fig:SEDevo}).  The forward shock also heats electrons to ultra-relativistic temperatures, powering (incoherent) synchrotron X-ray/gamma-ray emission, similar to a gamma-ray burst afterglow ($\S\ref{sec:afterglow}$; Fig.~\ref{fig:afterglow}).  On larger scales, the train of ion shells from consecutive flares merges into a wind that feeds the nebula through a termination shock.  Electrons injected at the termination shock powers the persistent radio source and (after cooling) generates the large rotation measure of the bursts.  Stochastic or secular variation in the burst DM can also arise from the ion shell (on timescales of $\Delta T \lesssim$ days) or from photo-ionization of the supernova ejecta by the flare X-rays (on timescales of the source age of years to decades).    }
\label{fig:schematic}
\end{figure*}

This paper is organized as follows.  In $\S\ref{sec:dynamics}$ we review the dynamics of shock deceleration of an ultra-relativistic flare of ejecta by a slowly expanding upstream medium (the upstream ion wind inferred for FRB 121102).  In $\S\ref{sec:FRB}$ we combine the dynamics with the result of PIC simulations to predict the time-dependent synchrotron maser emission, which we apply to the above questions, particularly motivated by observations of FRB 121102; an important ingredient in the observed emission is the role of induced scattering ($\S\ref{sec:scattering}$). In $\S\ref{sec:afterglow}$ we describe the coincident synchrotron afterglow of the flares.  In $\S\ref{sec:discussion}$ we summarize our results and expand on their implications.

\section{Shock Deceleration of the Flare Ejecta}
\label{sec:dynamics}

Consider a scenario in which a central engine suddenly injects an isotropic energy $E$ over a short duration $\delta t \lesssim 10^{-4}-10^{-3}$ s, producing a radially-expanding shell with an initial bulk Lorentz factor $\Gamma_{\rm ej} \gg 1$.  This ultra-relativistic shell collides with a sub-relativistic (effectively stationary) external medium characterized by a power-law radial density profile $n_{\rm ext} \propto r^{-k}$, where $k < 3$.   

This section reviews how the ultra-relativistic shell transfers the energy $E$ to the surrounding medium through a time-dependent shock wave.  While our description can be generalized to any central engine, in $\S\ref{sec:upstream}$ we first review order-of-magnitude estimates for the properties of the fast ejecta and external medium in the flaring magnetar scenario, as these will motivate numerical evaluation of key expressions used later.

\subsection{Upstream Ion Wind in Flaring Magnetar Scenario}
\label{sec:upstream}

A neutron star born with a strong internal magnetic field $\gtrsim 10^{16}$ G, possibly as a result of rapid rotation at birth \citep{Duncan&Thompson92}, possesses a total reservoir of magnetic energy of $E_{B_{\star}} \gtrsim 10^{50}$ erg.  The magnetic field is predicted to leak out of the star over a timescale $t_{\rm life} \sim 10-100$ yr set by ambipolar diffusion in its core \citep{Beloborodov&Li16}, similar to the inferred age of the source responsible for FRB 121102 (e.g.~\citealt{Metzger+17}).  The emergence of magnetic energy is unlikely to be a steady process, but instead could occur in discrete bursts perhaps similar to the flares from significantly older and less active magnetars in our Galaxy.

\newpage

\citet{Opperman+18} found a mean repetition rate of 5.7$^{+3.0}_{-2.0}$ bursts per day for FRB 121102, corresponding to an average interval $\Delta T \sim 10^{4}$ s.  The repetition pattern is non-Poissonian \citep{Opperman+18}, indicating that the bursts are often clustered in time such that $\Delta T$ can be substantially shorter (e.g. 6 of the 11 bursts from \citet{Spitler+16} were detected within a 10 minute period), with median intervals between flares of hundreds of seconds (see also \citealt{Katz18,Li+19} for detailed analysis).  However, weighted by radiated energy, the luminosity function of FRB 121102 is dominated by the rare highest fluence bursts, which take place at a rate $\lesssim 1$ day$^{-1}$ ($\Delta T \gtrsim 10^{5}$ s; e.g.~\citealt{Nicholl+17,Law+17}).  The total energy available between each strong flare is then 
\be E_{\rm tot} \sim (E_{B_{\star}}/t_{\rm life})\Delta T \sim 10^{45}-10^{46}{\rm erg}. \ee  

In our scenario, this energy is shared between at least one "clean" initial ultra-relativistic $\Gamma_{\rm ej} \gg 1$, potentially highly-magnetized $\sigma \gg 1$ pulse of energy $E$ at the beginning of the flare responsible for the powering the FRB \citep{Lyubarsky14,Beloborodov17} and a more prolonged phase of ion-loaded mass-loss which emerges with a sub-relativistic velocity $\beta_w = v_w/c \lesssim 1$ and lower magnetization $\sigma \lesssim 1$.  The latter forms the upstream medium into which the clean pulse from subsequent flares collides, as well as feeds the nebula electrons to power the persistent radio source and generate its high RM \citep{Margalit&Metzger18}.

While the physical mechanism, and thus the time-dependence, of the ion mass loss is theoretically uncertain, for FRB 121102 the time-averaged ion injection rate $\dot{M}$ at the present epoch is constrained to be $\sim 10^{19}-10^{21}$ g s$^{-1}$ (\citealt{Margalit&Metzger18}).  The kinetic energy carried by the ion ejecta of each major flare,
\be E_w \sim (\dot{M}v_{\rm w}^{2}/2)\Delta T \sim 10^{46}\,{\rm erg}\,\,\dot{M}_{21}\Delta T_{5}\left(\frac{\beta_w}{0.5}\right)^{2}, \label{eq:Edotw}
\ee
is therefore within the magnetar's budget, $\sim E_{\rm tot}$ \citep{Margalit&Metzger18}.  Here and hereafter we employ the short-hand notation $q_x = q/10^{x}$ in cgs units, e.g. $\dot{M}_{21} = \dot{M}/(10^{21}$ g s$^{-1})$.  

We consider two limits for the time-dependence of the ion-electron wind.  First, if the ions were to emerge from the magnetar isotropically at a strictly constant rate, then the radial density profile would be that of a steady wind, 
\be
n_{\rm ext} = \frac{\dot{M}}{4\pi v_w r^{2} m_p} \,\,\,\,\text{steady\,wind\,($k = 2$).}
\label{eq:nwind}
\ee 

Perhaps more realistically, the ions are released in temporally-concentrated episodes following each major flare.  This scenario is consistent with the high mass-loading of the ejecta inferred from the radio afterglow of the 2004 giant flare from SGR 1806-20 \citep{Gelfand+05,Taylor+05,Granot+06}.  On average, the ion shell from each flare must contain sufficient mass, $\Delta M  = \dot{M}\Delta T$, to produce the same time-averaged value of $\dot{M}$.  At large radii, $r \gg r_{\rm s}$, where 
\be r_{\rm s} \equiv v_w \Delta T \sim 1.5\times 10^{15}{\rm cm}\,\left(\frac{\beta_w}{0.5}\right) \Delta T_{5},
\label{eq:rs}
\ee
the train of ion shells from consecutive flares will merge to form a steady-wind of density similar to equation (\ref{eq:nwind}).  

However, at radii $r \ll r_{\rm s}$ the external density encountered by the next ultra-relativistic shell will be much smaller than for a steady wind (eq.~\ref{eq:nwind}).  While not zero, because the trans-relativistically expanding ion shell has time to spread radially due to finite dispersion in its velocity, the density power-law slope will be much shallower than $\propto r^{-2}$.  

Under the assumption that $n_{\rm ext}$ is radially constant for $r \ll r_{\rm s}$, the external density profile at radii $r \ll r_{\rm s}$ by the time of the next strong flare will be given by
\be
n_{\rm ext}(r \ll r_{\rm s}) = \frac{3\Delta M}{4\pi m_p r_{\rm s}^{3}} \approx \frac{3\dot{M}\Delta T}{4\pi m_p r_{\rm s}^{3}}\,\,\,\,\text{discrete\,shell \,($k = 0$)}.
\label{eq:ndiscrete}
\ee
While we have chosen a constant density profile somewhat arbitrary, our qualitative conclusions to follow for similarly "flat" profiles (as compared to the steady-wind case) are robust to this detail.
See Figure \ref{fig:schematic} for a schematic illustration.

\subsection{Dynamics of the Shell Deceleration}

The deceleration of the ejecta shell by the external medium, and its resulting radiation, occurs in two phases.  During the initial phase a reverse shock crosses back through the ejecta shell and the ejecta energy is transferred to the forward shock (\citealt{Sari&Piran95}).\footnote{ If the ultra-relativistic ejecta shell is highly magnetized, the reverse shock will be weak; however, the dynamics of the forward shock, of greatest interest here as the likely site of the FRB emission (see $\S\ref{sec:FRB}$), are relatively insensitive to the dynamics of the reverse shock.
}  This process completes once the reverse shock passes entirely through the ejecta shell, as occurs at the deceleration radius, $r_{\rm dec}$.  This phase is then followed by a self-similar deceleration of the forward shock at radii $r \gg r_{\rm dec}$ (\citealt{Blandford&McKee76}).  Although this overall evolution is well documented in the GRB literature (e.g.~\citealt{Kumar&Zhang15}), we repeat it here for purposes of clarity.

\subsubsection{Early and Late Deceleration Phases}

Consider first the early deceleration phase ($r \ll r_{\rm dec}$).  In the rest frame of the upstream medium, the ultra-relativistic shell of thickness $\Delta = c \cdot \delta t$ does not have time to expand radially as it moves outwards into the external medium.  The co-moving density in the unshocked ejecta shell at radius $r$ is thus given by
\be
n_{\rm ej} \simeq \frac{E}{\delta t}\frac{1}{4\pi r^{2}  m_p c^{3} \Gamma_{\rm ej}^{2}},
\label{eq:nej}
\ee
where we have assumed a cold ejecta shell dominated by its bulk kinetic energy.  The ratio of the density of the ultra-relativistic ejecta shell (eq.~\ref{eq:nej}) to that of the external medium $n_{\rm ext} \propto r^{-k}$ is defined as
\be
f \equiv \frac{n_{\rm ej}}{n_{\rm ext}} \propto r^{k-2}.
\ee
When $f \gg 1/\Gamma_{\rm ej}^{2}$, as in all cases of present interest, the reverse shock is ultra-relativistic \citep{Sari&Piran95}.  It crosses the ejecta shell on a timescale and by a radius given, respectively, by\footnote{The pre-factor here can vary moderately from the assumed value of 2, depending on the details of the hydrodynamical evolution (e.g.~\citealt{Sari97}).  We nevertheless adopt this factor to follow common convention.}
\be
\tilde{t}_{\rm dec} \approx 2\Gamma^{2}(t_{\rm dec})\delta t,\,\,\, r_{\rm dec} = c\tilde{t}_{\rm dec},
\ee
where a tilde denotes time in the rest-frame of the upstream medium, which we approximate as being stationary ($\beta_w \ll 1$), and $\Gamma$ is the Lorentz factor of the shocked gas.  The latter obeys \citep{Sari&Piran95}
\be
\Gamma (r \ll r_{\rm dec}) = \left(\frac{f\Gamma_{\rm ej}^{2}}{4}\right)^{1/4} \propto r^{\frac{(k-2)}{4}} = \left\{
\begin{array}{lr} 
r^{-1/2} & k = 0  \\
r^{0} & k = 2  \\ 
\end{array}
\right. . 
\label{eq:RC}
\ee
during the reverse shock crossing phase.

At times $\tilde{t} \gg \tilde{t}_{\rm dec}$, or radii $r \gg r_{\rm dec}$, the forward shock evolution approaches the \citet{Blandford&McKee76} self-similar form,
\begin{eqnarray}
&&\Gamma (r \gg r_{\rm dec}) \nonumber \\
 & =& \left(\frac{17-4k}{16\pi}\frac{E}{m_p n_{\rm ext} r^{3}c^{2}}\right)^{1/2} \propto r^{\frac{(k-3)}{2}} = \left\{
\begin{array}{lr} 
r^{-3/2}, & k = 0  \\
r^{-1/2}, & k = 2  \\
\end{array}
\right. .  \nonumber \\
\label{eq:BM}
\end{eqnarray}
The transition in the evolution of $\Gamma (t)$ at $r_{\rm dec}$ between that given by equations \ref{eq:RC} and \ref{eq:BM} is smooth,
\be
\Gamma (r \gg r_{\rm dec}) \approx \Gamma (r_{\rm dec})\left(\frac{r}{r_{\rm dec}}\right)^{\frac{(k-3)}{2}}.
\ee
The above relations assume adiabatic (energy conserving) evolution of the shock.  This is justified because, although electrons cool efficiently through synchrotron radiation ($\S\ref{sec:afterglow}$), ions$-$which likely hold the majority of the energy$-$do not.

\subsubsection{Full Time Evolution}
\label{sec:evo}

We now summarize various properties of the shock as a function of time $t \simeq (r/c)(1-\beta) \simeq r/(2c\Gamma^{2}$) as measured by an observer ahead of the shock, for both the early and late deceleration phases.  The luminosity of the shock as seen by an observer directly ahead of the shock (within its $1/\Gamma$ cone) is given by
\be
L_{\rm sh} \approx 4\pi r^{2} n_{\rm ext}\Gamma ^{4} m_p c^{3} \propto \Gamma ^{4}r^{2-k}.
\label{eq:Lsh}
\ee

First note the observed deceleration time equals the duration of central engine activity,
\be
t_{\rm dec} \approx \frac{\tilde{t}_{\rm dec}}{2\Gamma ^{2}} \sim \delta t,
\ee
a well-known result from GRBs.  At earlier times $t \ll t_{\rm dec} \sim \delta t$, 
\begin{eqnarray}
r \propto t^{\frac{2}{(4-k)}} &=& \left\{
\begin{array}{lr} 
t^{1/2}, & k = 0  \\
t^{1}, & k = 2 \\
\end{array}
\right. ,
\end{eqnarray}
\begin{eqnarray}
 \Gamma  \propto t^{\frac{(k-2)}{2(4-k)}} &=& \left\{
\begin{array}{lr} 
t^{-1/4}, & k = 0  \\
t^{0}, & k = 2 \\
\end{array}
\right. ,
\end{eqnarray}
\begin{eqnarray}
n_{\rm ext}(r) \propto t^{\frac{-2k}{(4-k)}} &=& \left\{
\begin{array}{lr} 
t^{0}, & k = 0  \\
t^{-2}, & k = 2 \\
\end{array}
\right. ,
\end{eqnarray}
\begin{eqnarray}
L_{\rm sh}  \propto n_{\rm ext}\Gamma ^{4}r^{2} \propto t^{0},\,\,\,\,\,\, k = 0, 2
\end{eqnarray}
For times $t \gg t_{\rm dec} \sim \delta t$ we have
\begin{eqnarray}
r_{\rm sh} \propto  t^{\frac{1}{(4-k)}} &=& \left\{
\begin{array}{lr} 
t^{1/4}, & k = 0  \\
t^{1/2}, & k = 2 \\
\end{array}
\right. ,
\end{eqnarray}
\begin{eqnarray}
\Gamma  \propto t^{\frac{(k-3)}{2(4-k)}}  &=& \left\{
\begin{array}{lr} 
t^{-3/8}, & k = 0  \\
t^{-1/4}, & k = 2 \\
\end{array}
\right. ,
\end{eqnarray}
\begin{eqnarray}
n_{\rm ext}(r_{\rm sh}) \propto t^{\frac{-k}{(4-k)}}  &=& \left\{
\begin{array}{lr} 
t^{0}, & k = 0  \\
t^{-1}, & k = 2 \\
\end{array}
\right. ,
\end{eqnarray}
\begin{eqnarray}
L_{\rm sh}  \propto n_{\rm ext}\Gamma ^{4}r_{\rm sh}^{2} \propto t^{-1}, \,\,\,\,\,\,k = 0, 2 \label{eq:Lext}
\end{eqnarray}
The fluence released by the shock $\sim L_{\rm sh}t$ per decade in time is therefore relatively constant at times $t \gg \delta t$.
As discussed below, these time-dependent properties give rise to time-evolving FRB emission (\S\ref{sec:FRB}) and a broad-band synchrotron afterglow (\S\ref{sec:afterglow}).

\subsubsection{Numerical Values in Magnetar Model}
\label{sec:numerical}

Here we provide numerical estimates of relevant shock quantities at $r = r_{\rm dec}$ ($t = \delta t$), separately for the steady wind ($k = 2$; eq.~\ref{eq:nwind}) and discrete shell ($k = 0$; eq.~\ref{eq:ndiscrete}) scenarios for the upstream medium.  Combined with the power-law evolution specified in \S \ref{sec:evo}, these determine their values at all times.

\paragraph{Steady Wind ($k = 2$).}
For a steady-wind external medium, the Lorentz factor of the shocked gas (eq.~\ref{eq:RC}) is constant during the early reverse shock-crossing phase,
\be 
\Gamma (r \ll r_{\rm dec})  = \left(\frac{E/\delta t}{4\dot{M}c^{2}}\beta_w\right)^{1/4} \approx 7.3 E_{43}^{1/4}\dot{M}_{21}^{-1/4} \beta_w^{1/4}\delta t_{-3}^{-1/4} ,
\ee
independent of both radius and the initial Lorentz factor $\Gamma_{\rm ej}$.  The deceleration radius is then 
\be
r_{\rm dec} \approx 2\Gamma ^{2}c\delta t  \approx 3.2\times 10^{9}{\rm cm}\, E_{43}^{1/2}\dot{M}_{21}^{-1/2} \beta_w^{1/2}\delta t_{-3}^{1/2}.
\label{eq:rdecwind}
\ee
The upstream density at the location of the shock is (eq.~\ref{eq:nwind})
\be
n_{\rm ext}(r_{\rm dec}) = \frac{\dot{M}}{4\pi r_{\rm dec}^{2} m_p v_w} \approx  6\times 10^{14}{\rm cm^{-3}} E_{43}^{-1}\dot{M}_{21}^{2}\left(\frac{\beta_w}{0.5}\right)^{-2}\delta t_{-3}^{-1}.
\ee
The shock luminosity (eq.~\ref{eq:Lsh}) at $t \lesssim t_{\rm dec}$ is constant,
\begin{eqnarray}
L_{\rm sh}(r_{\rm dec}) \approx \frac{E}{4\delta t} \approx 2.5\times 10^{45}{\rm erg\,s^{-1}}E_{43}\delta t_{-3}^{-1}. \label{eq:Lshconst}
\end{eqnarray}
The optical depth to Thomson scattering ahead of the shock is
\be
\tau_{\rm T}(r_{\rm dec}) = \frac{\dot{M}\kappa_{\rm es}}{4\pi v_w r_{\rm dec}} \approx 0.5 E_{43}^{-1/2}\dot{M}_{21}^{3/2}\left(\frac{\beta_w}{0.5}\right)^{-3/2}\delta t_{-3}^{-1/2},
\label{eq:tauTw}
\ee
where we have taken $\kappa_{\rm es} \approx 0.2$ cm$^{2}$ g$^{-1}$ for an assumed electron/heavy ion composition.  The DM ahead of the shock is given by
\begin{eqnarray}
{\rm DM} &=& \int \left(\frac{n_{\rm ext}}{2}\right)dr \simeq \frac{\dot{M}}{8\pi m_p v_w r_{\rm dec}}  \nonumber \\
&\approx& 2\times 10^{5}\,{\rm pc\,cm^{-3}} E_{43}^{-1/2}\dot{M}_{21}^{3/2}\left(\frac{\beta_w}{0.5}\right)^{-3/2}\delta t_{-3}^{-1/2}.
\label{eq:DMwind}
\end{eqnarray}

\paragraph{Discrete Ejecta Shell ($k = 0$).} 
For the case of the upstream medium being an ejecta shell from a previous flare (eq.~\ref{eq:ndiscrete}),
\begin{eqnarray}
\Gamma (r \ll r_{\rm dec}) &=& \left(\frac{E\beta_w}{12\dot{M} c^{2}\delta t}\right)^{1/4}\left(\frac{r}{r_{\rm s}}\right)^{-1/2} \nonumber \\
&\approx& 5.5E_{43}^{1/4}\dot{M}_{21}^{-1/4}\beta_w^{1/4}\delta t_{-3}^{-1/4}\left(\frac{r}{r_{\rm s}}\right)^{-1/2},
\end{eqnarray}
where the deceleration radius
\be
\frac{r_{\rm dec}}{r_{\rm s}} = 7.8\times 10^{-4}E_{43}^{1/4}\dot{M}_{21}^{-1/4}\beta_w^{-1/4}\Delta T_{5}^{-1/2}\delta t_{-3}^{1/4},
\ee
is a small fraction of the shell radius $r_{\rm s}$ (eq.~\ref{eq:rs}), consistent with our assumed density profile (eq.~\ref{eq:ndiscrete}).  In physical units,
\be
r_{\rm dec} \approx 2.3\times 10^{12}{\rm cm}\, E_{43}^{1/4}\dot{M}_{21}^{-1/4}\beta_w^{3/4}\Delta T_{5}^{1/2}\delta t_{-3}^{1/4}.
\label{eq:rdecdiscrete}
\ee
Note that, on timescales of milliseconds, the flare ejecta interacts with only a small fraction $\sim (r_{\rm dec}/r_{\rm s})^{3}$ of the shell mass.  

Solving for the Lorentz factor at the deceleration radius,
\begin{eqnarray}
\Gamma (r_{\rm dec}) \approx 196E_{43}^{1/8}\dot{M}_{21}^{-1/8}\beta_w^{3/8}\Delta T_{5}^{1/4}\delta t_{-3}^{-3/8}.
\end{eqnarray}
The shock luminosity is the same as in the wind case (eq.~\ref{eq:Lshconst}).  The density at the deceleration radius is (eq.~\ref{eq:ndiscrete})
\begin{eqnarray}
n_{\rm ext}(r_{\rm dec}) \approx 4\times 10^{3}{\rm cm^{-3}}\dot{M}_{21}\left(\frac{\beta_w}{0.5}\right)^{-3}\Delta T_{5}^{-2}.
\label{eq:ndiscrete2}
\end{eqnarray}

The total column ahead of the shock is now dominated by the mean radius of the shell $r_{\rm s} = v_{w}\Delta T$ instead of $r_{\rm dec}$, such that the Thomson optical depth ahead of the shock is
\be
\tau_{\rm T} = \frac{\dot{M}\kappa_{\rm es}}{4\pi v_w r_{\rm s}} \approx 7\times 10^{-7} \dot{M}_{21} \left(\frac{\beta_w}{0.5}\right)^{-2}\Delta T_{5}^{-1},
\label{eq:tauTdiscrete}
\ee
independent of $r_{\rm dec}$.  The local DM ahead of the shock is given by  
\be
{\rm DM} \simeq \frac{\dot{M}}{8\pi m_p v_w r_{\rm s}}  \approx 0.36\,{\rm pc\,cm^{-3}}\dot{M}_{21} \left(\frac{\beta_w}{0.5}\right)^{-2}\Delta T_{5}^{-1}.
\label{eq:DMdiscrete}
\ee

\section{Synchrotron Maser Emission (FRB)}
\label{sec:FRB}

\begin{figure}
\centering
\includegraphics[width=0.5\textwidth]{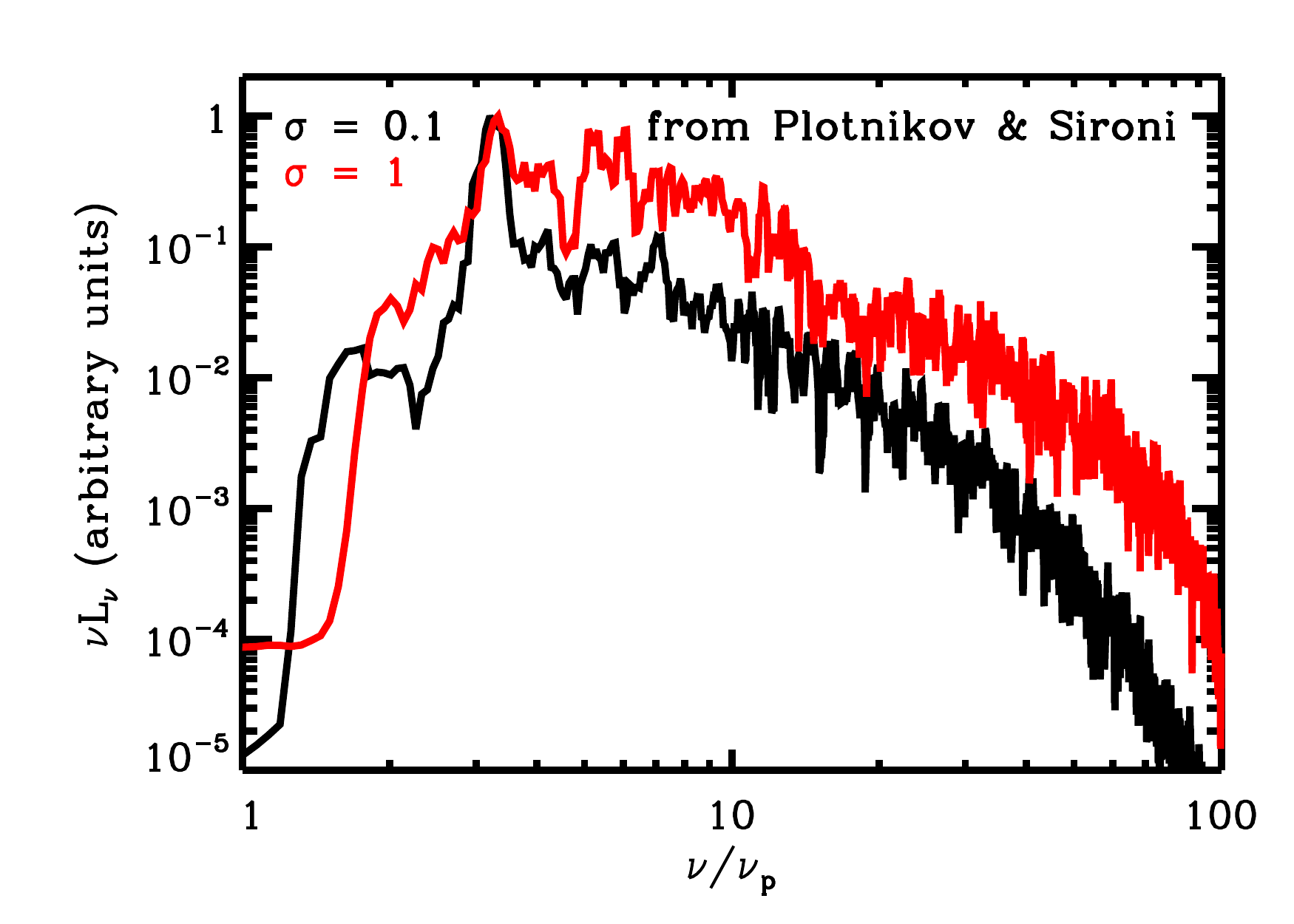}
\caption{Spectral energy distribution of FRB emission in the shock-powered synchrotron maser scenario, as calculated from PIC simulations by \citet{Plotnikov&Sironi19}, shown as a function of frequency in the rest-frame of emitting plasma, normalized to the upstream plasma frequency $\nu_{\rm p}$, which is close to the minimum cut-off frequency of the emission for low $\sigma$ (eq.~\ref{eq:numin}).  Two cases are shown for different values of the upstream magnetization, $\sigma = 0.1$ (black) and $\sigma = 1$ (red).  The two spectra are independently rescaled (in reality, the emission $\sigma=0.1$ is more luminous for fixed shock power than for $\sigma=1$). }
\label{fig:SED}
\end{figure}

Analysis of PIC simulations by \citet{Plotnikov&Sironi19} shows that the synchrotron maser produces an electromagnetic wave ahead of the shock with relatively narrowly peaked spectral energy distribution (SED), centered about the peak frequency
\begin{eqnarray}
\nu_{\rm pk} &\approx& \frac{1}{2\pi}(3\Gamma  \omega_p),  \label{eq:nuFRB}
\end{eqnarray}
where the factor $\Gamma$ accounts for the relativistic Doppler shift from the frame of the post-shock gas into that of the observer (we neglect cosmological redshift effects) and the precise prefactor varies moderately in the range $\sigma \sim 0.1-1$ of interest.  Here $\omega_p = (4\pi n_e e^{2}/m_e)^{1/2}$ is the plasma frequency of the medium ahead of the shock and $n_e$ is the electron density of the upstream medium. Figure \ref{fig:SED} shows the predicted SED from \citet{Plotnikov&Sironi19} for two values of the upstream magnetization, $\sigma = 0.1$ and $\sigma= 1$.  

A key feature of the SED is the presence of a cut-off at a minimum frequency, 
\be \nu_{\rm min} \simeq  \frac{1}{2\pi}(\Gamma  \omega_p) \approx \frac{\nu_{\rm pk}}{3}
\label{eq:numin}
\ee 
below which the shock front out-runs the precursor FRB \citep{Plotnikov&Sironi19}. While the lower frequency cut-off is sharp, the SED contains significant power even at frequencies $\gg \nu_{\rm pk}$ contributed by higher-order harmonics of the synchrotron maser.  

One caveat of applying these results in the present context is that the calculations assumed an upstream medium of electron/positron composition; for an ion-electron plasma the SED could in principle be different.  Although the bulk of the energy carried away from the FRB 121102 source is inferred to be an ion-electron plasma \citep{Margalit&Metzger18}, the FRB is produced by a shock that passes through only a small fraction of this material, representing either the very tail-end of the flare ejecta or that injected intermittently between the major flares by the rotationally-powered component of the magnetar wind, which could have a different (e.g.~ electron/positron) composition than the bulk.  Furthermore, it is not clear whether the maser emission from electron-ion shocks \citep{Hoshino08,Sironi&Spitkovsky10}  will qualitatively differ from the electron/positron case.

As we now discuss, the observed FRB emission is more likely to originate from the forward shock into the upstream wind than the reverse shock back into the ejecta.  One piece of evidence comes from the nearly constant polarization angle of FRB 121102 over the course of many bursts \citep{Michilli+18}.  Given the complex nature of magnetar flares, it would be surprising if the orientation of the magnetic field in the ultra-relativistic ejecta$-$to which the polarization angle of the maser emission is perpendicular$-$were similar for each flare.  By contrast, the magnetic field carried by the slower ion-electron wind, which likely emerges from the magnetar surface over many rotation periods, could be more easily shaped into an orientation perpendicular to its fixed rotation axis.  The electron-ion external medium is also likely to possess a lower magnetization than the cleaner flare ejecta, which also favors the forward shock because of the decreasing efficiency of the coherent emission for higher $\sigma$, especially given the strong $1/\sigma^{2}$ scaling at $\sigma\gg 1$ \citep{Plotnikov&Sironi19}.  Finally, FRB emission from the reverse shock would need to pass through relativistically hot gas, which could attenuate the signal due to induced Compton scattering \citep{Lyubarsky08}.

\begin{figure}
\includegraphics[width=0.5\textwidth]{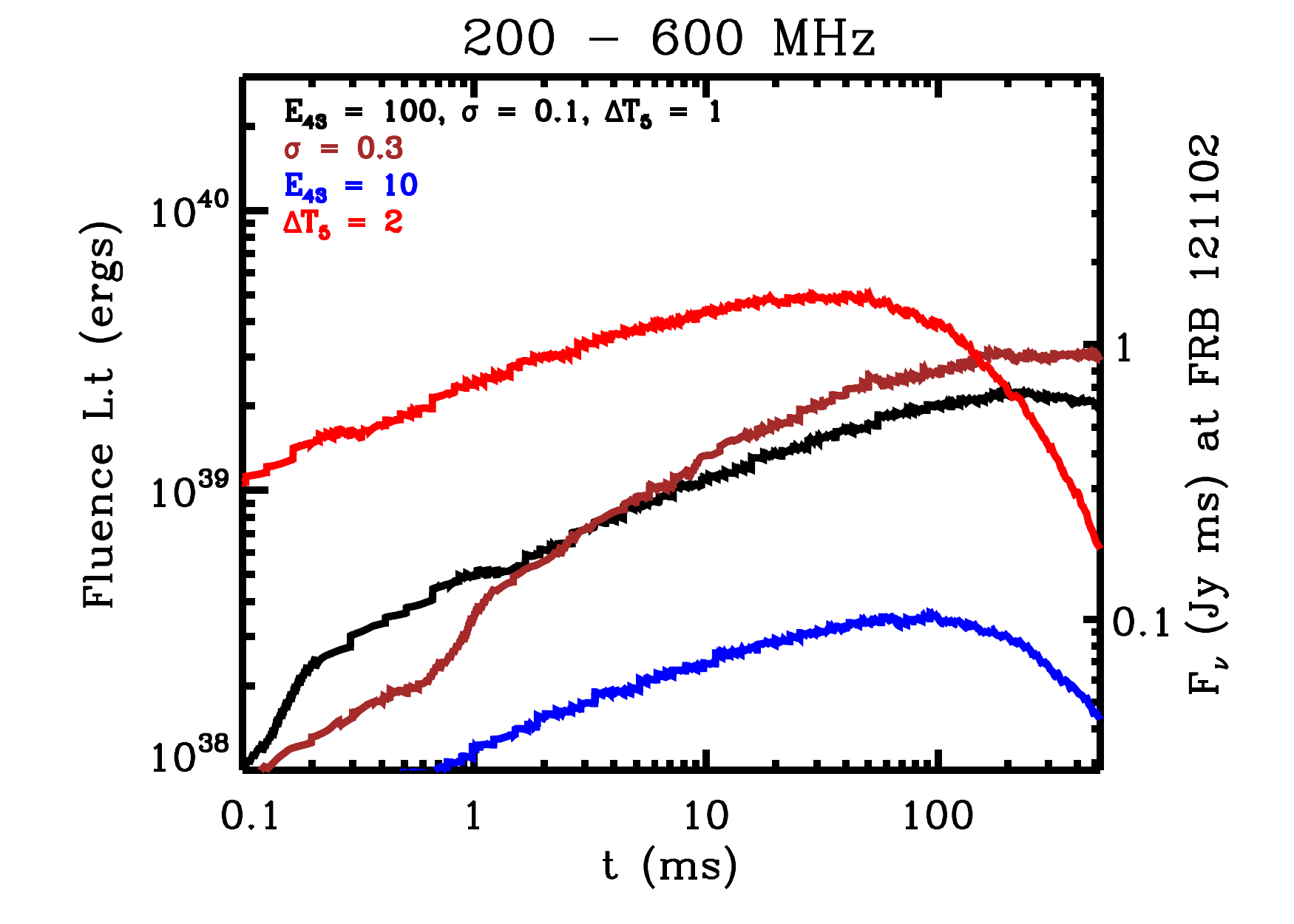}
\includegraphics[width=0.5\textwidth]{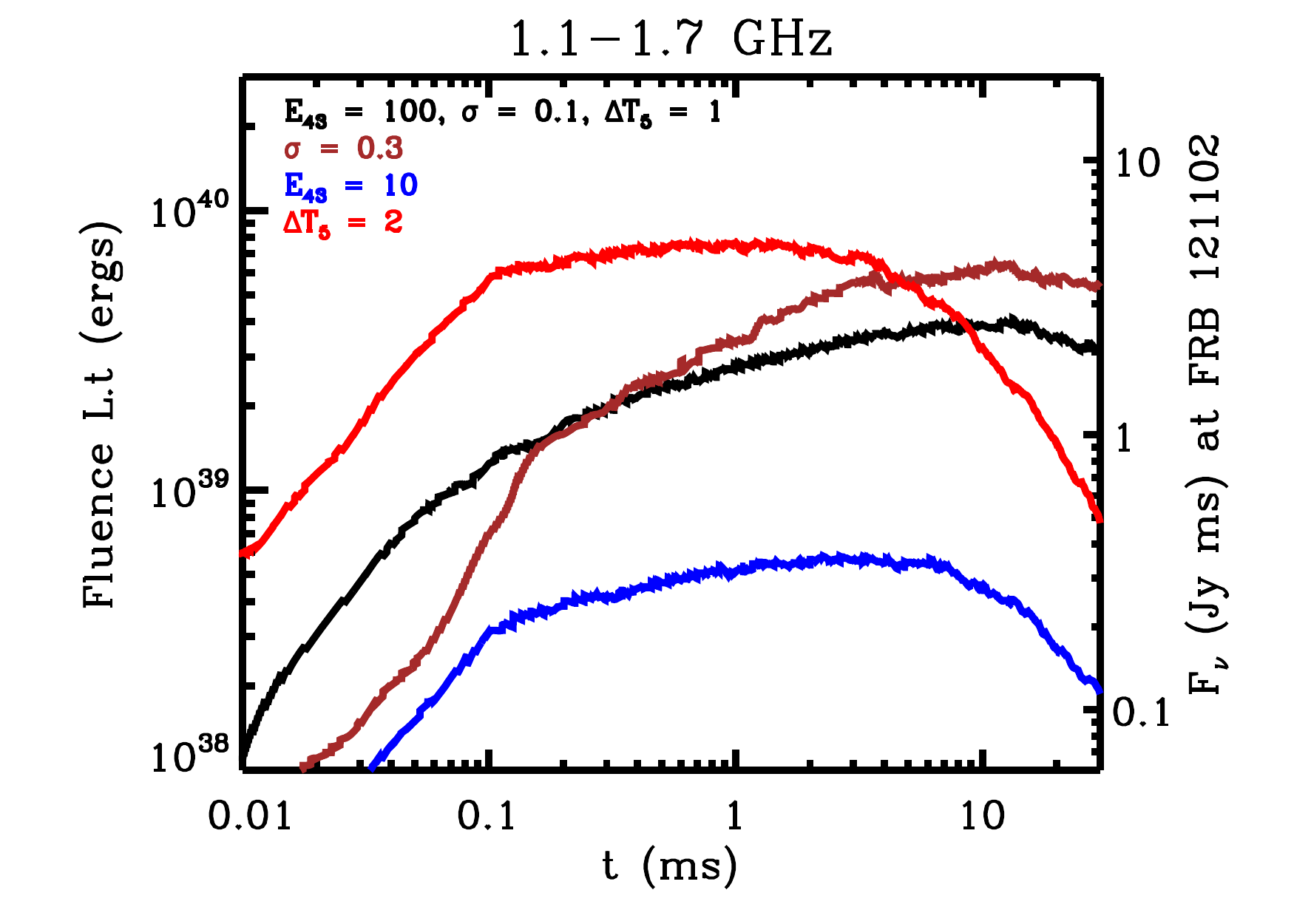}
\includegraphics[width=0.5\textwidth]{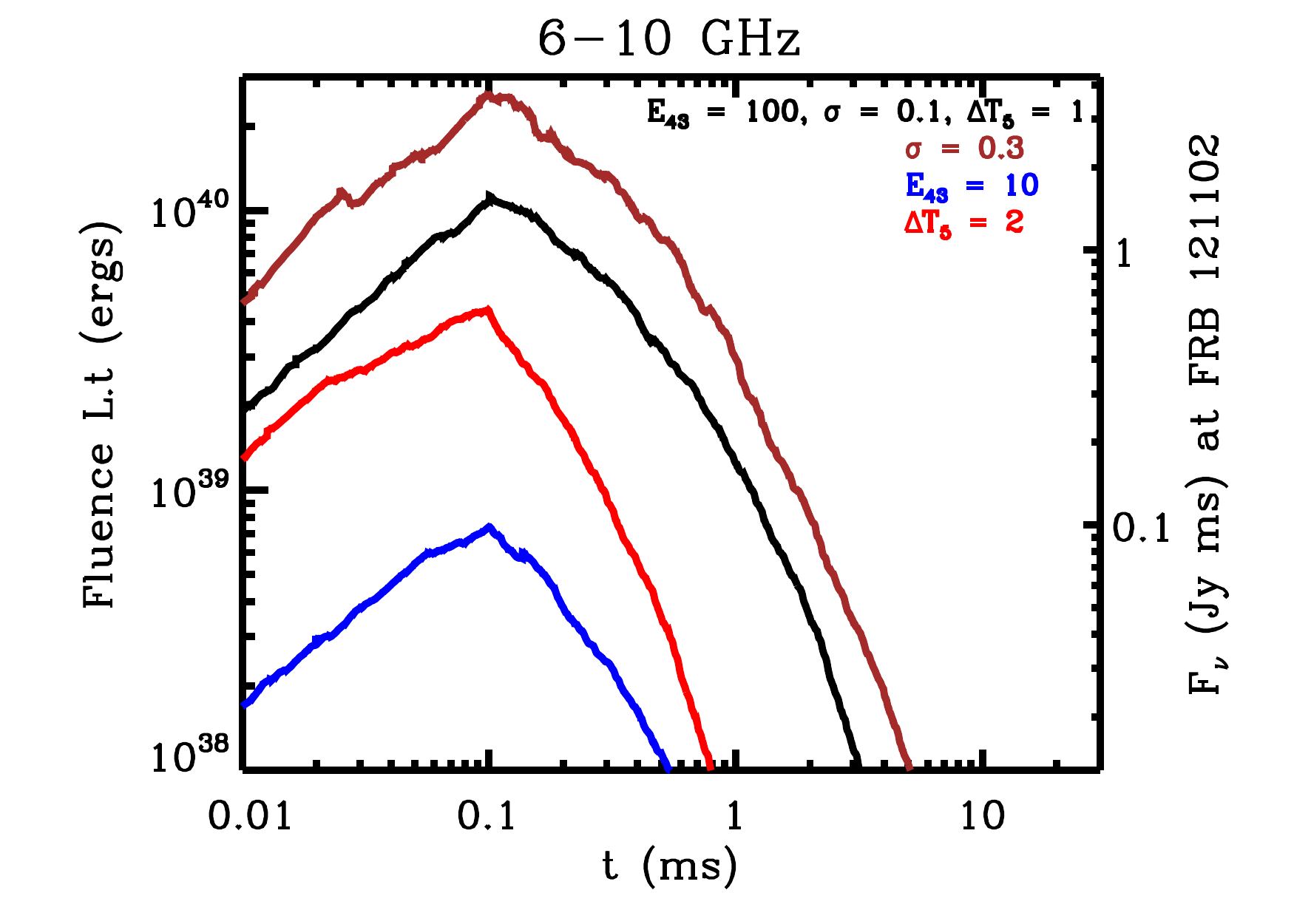}
\caption{Theoretical FRB light curves, expressed as fluence $t \cdot \int L_{\nu} d\nu$ as a function of time $t$ over the 0.2$-$0.4 GHz (top), 1.1$-$1.7 GHz (middle), and 6$-$10 GHz (bottom) spectral bands.  These are calculated by combining the time-dependent FRB luminosity and peak frequency of the decelerating blast-wave (constant density external medium $k = 0$) with the predicted SED (Fig.~\ref{fig:SED}), accounting for attenuation at early times due to induced scattering according to the effective optical depth given by equation (\ref{eq:tauC3}).  Different colors show models calculated for different assumptions about the magnetization of the upstream medium, $\sigma$.  The parameters of the baseline model (black line) are: $E = 10^{45}$ erg; $\Delta T = 10^{5}$ s; $\delta t = 10^{-4}$ s, $\beta_w = 0.5$, $f_{\xi} = 10^{-3}$, $\dot{M}_{21} = 1$.  Colors show models with the values of $\sigma$, $E$, and $\Delta T$ varied about the fiducial models as marked.
}
\label{fig:lightcurves}
\end{figure}

\begin{figure}
\centering
\includegraphics[width=0.5\textwidth]{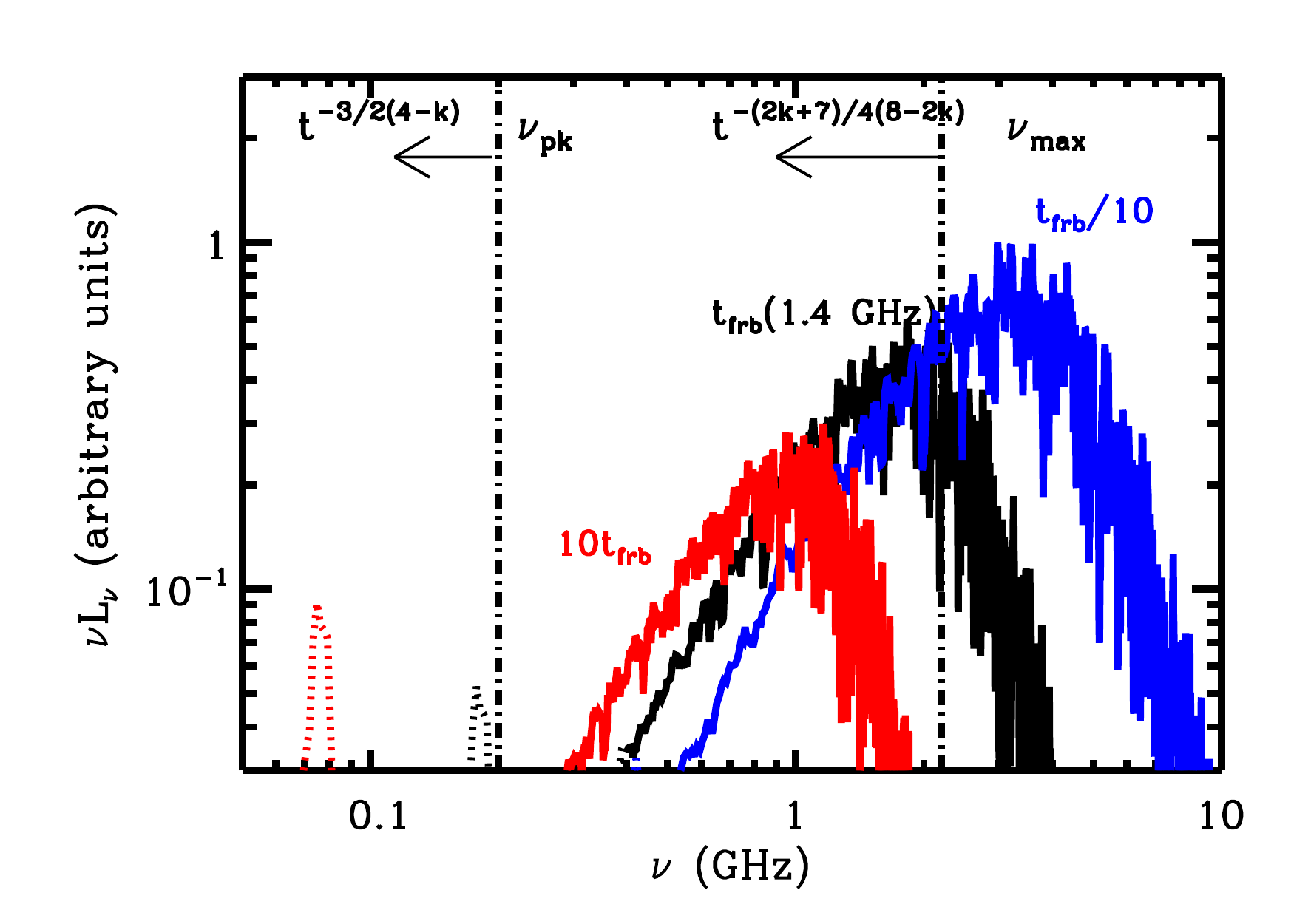}
\caption{Spectral energy distribution of radio emission escaping from the vicinity of the shock at three snapshots in time relative to the 1.4 GHz duration $t_{\rm frb}$(1.4 GHz) for a model with $\delta t = 10^{-4}$ s, $E = 10^{44}$ erg, $\Delta T = 10^{5}$ s, $\beta_w = 0.5$, $f_{\xi} = 10^{-3}$, $\sigma = 0.1$.  As the blast wave decelerates, the intrinsic frequency of the maser emission drifts downward at a rate $\nu_{\rm pk} \propto t^{-3/(8-2k)}$ (eq.~\ref{eq:nuFRBevo}) which depends on the power-law index of the upstream medium density profile, $n_{\rm ext} \propto r^{-k}$.  Some radiation may escape just below this frequency (the spikes that appear at low frequencies) but as the fluence is highly uncertain, we have denoted this part of the spectrum with a dotted lines.  The suppression of the SED across intermediate frequencies is the result of induced Compton scattering, as accounted for approximately by the effective optical depth $\tau_{\rm c}$ given by equation (\ref{eq:tauC3}).  Most of the fluence escapes near the critical frequency, $\nu_{\rm max}$, above which $\tau_{\rm c} \lesssim 3$, which drifts downward as a more gradual power-law of time, $\nu_{\rm max} \propto t^{-(2k+7)/4(8-2k)}$ for $t \gtrsim \delta t$ (eq.~\ref{eq:numax2}).  Due to the sensitive dependence of $\tau_{\rm c} \propto (d/d\nu)(L_{\nu}/\nu)$ on the spectral slope, in a more detailed treatment the spectral maximum is likely to be sharpened.}
\label{fig:SEDevo}
\end{figure}

Assuming FRB emission originates from the forward shock, then using results from \S\ref{sec:evo}, we find that the peak frequency of the FRB pulse at the forward shock evolves during the early deceleration phase ($t \lesssim \delta t$) as
\begin{eqnarray}
\nu_{\rm pk} \propto n_{\rm ext}^{1/2}\Gamma  \propto t^{-\frac{(k+2)}{8-2k}}= \left\{
\begin{array}{lr} 
t^{-1/4} & k = 0  \\
t^{-1} & k = 2 \\
\end{array}
\right. ,
\end{eqnarray}
while during the later deceleration phase ($t \gtrsim \delta t$) we have
\begin{eqnarray}
\nu_{\rm pk} \propto t^{-\frac{3}{8-2k}} = \left\{
\begin{array}{lr} 
t^{-3/8} & k = 0  \\
t^{-3/4} & k = 2 \\
\end{array}
\right. .
\label{eq:nuFRBevo}
\end{eqnarray}

Using numerical values from \S\ref{sec:numerical} for $\Gamma$ and taking $n_e \approx n_{\rm ext}/2$ for our assumed electron-heavy ion composition, we obtain (eq.~\ref{eq:nuFRB}) 
\begin{eqnarray}
\nu_{\rm pk}(t > t_{\rm dec}) \approx \left\{
\begin{array}{lr} 
0.18\,{\rm GHz}\,\,\ E_{43}^{1/8}\dot{M}_{21}^{3/8}\left(\frac{\beta_w}{0.5}\right)^{-9/8}\Delta T_{5}^{-3/4}t_{-3}^{-3/8} & k = 0 \\
2.9\times 10^{3}\,{\rm GHz}\,\,\,E_{43}^{-1/4}\left(\frac{\beta_{w}}{0.5}\right)^{-3/4}\dot{M}_{21}^{3/4}t_{-3}^{-3/4} & k = 2  \\

\end{array} 
\label{eq:nuFRBevo2}
\right. \nonumber \\
\end{eqnarray}
The peak frequency of the maser emission, and thus intrinsic structure in the SED, will {\it decrease in time} as the ultra-relativistic ejecta decelerates.  

If the electromagnetic wave created by the synchrotron maser carries a fraction $f_{\xi} = 10^{-3}f_{\xi,-3}$ of the luminosity of the forward shock (eq.~\ref{eq:Lsh}), then the predicted evolution of the bolometric luminosity of the FRB at times $t \gg t_{\rm dec}$ is given (in both steady-wind and discrete shell scenarios) by
\begin{eqnarray}
\nu L_{\nu}|_{\nu_{\rm pk}}(t \gtrsim t_{\rm dec}) \approx  f_{\xi}\frac{E}{4t} \approx 3\times 10^{42}\,{\rm erg\,s^{-1}}f_{\xi,-3}E_{43}t_{-3}^{-1}
\label{eq:Lnu}
\end{eqnarray}
The radiative efficiency depends on the upstream magnetization, with 1D PIC simulations predicting
$f_{\xi} \sim 0.03$ for $\sigma = 0.1-0.4$ and $f_{\xi} \approx 7\times 10^{-4}\sigma^{-2}$ for $\sigma \gg 1$ \citep{Plotnikov&Sironi19}.  Based on the persistent emission from the nebula surrounding FRB 121102 and its effects on FRB propagation, the ion-loaded outflow need be only moderately magnetized, e.g.~$\sigma \sim 0.1-0.5$ (e.g.~\citealt{Vedantham&Ravi18,Gruzinov&Levin19}), in which case $f_{\xi} \sim 0.003-0.03$.  On the other hand, multi-dimensional effects, or the presence of ions, could reduce the efficiency predicted from 1D electron/positron models by a factor of $10$ in the case of low $\sigma$ (\citealt{Iwamoto+17,Iwamoto+18}; Sironi et al., in prep).  The FRB efficiency of the shock can also be suppressed if the upstream medium is relativistically hot (Babul et al., in prep).  Although gamma-rays from the shock will heat the upstream medium via Compton scattering, we show in $\S\ref{sec:afterglow}$ that the temperatures achieved are generally not sufficiently high to reduce the value of $f_{\xi}$.

\subsection{Induced Scattering}
\label{sec:scattering}

An important general constraint on the site of FRB emission comes from potential suppression of the short radio pulse due to induced large-angle scattering of radially-directed rays by electrons in the upstream medium through the Compton and Raman processes  \citep{Lyubarsky08}.  The effective optical depth for induced Compton scattering of an electromagnetic pulse of frequency $\nu$,  luminosity $L_{\nu}$, and duration $t$ passing through a medium of electron density $n_e = n_{\rm ext}/2$ and radius $r$ from the central source is estimated by (\citealt{Lyubarsky08,Lyubarsky&Ostrovska16})
\begin{eqnarray}
\tau_{\rm c} &\approx& \frac{1}{10}\left(\frac{3}{64 \pi^{2}}\frac{\sigma_T}{m_e}\frac{ct n_{\rm ext}}{r^{2}}\right)\frac{\partial}{\partial \nu}\left(\frac{L_{\nu}}{\nu}\right),
\label{eq:tauC}
\end{eqnarray}
where the prefactor of $1/10$ is the suggested threshold for substantial attenuation by \citet{Lyubarsky08}, based on the additional time required for radiation at large angles to the primary beam to grow from its low initial background level.  

Equation \ref{eq:tauC} shows that scattering requires the photon spectrum of the primary beam $L_{\nu}/\nu$ to have a positive slope.  Initially, this condition is satisfied by the narrowly-peaked synchrotron maser SED only below its spectral peak at $\nu \sim \nu_{\rm pk}$ (Fig.~\ref{fig:SED}).  Making the approximation that $(\partial/\partial \nu)(L_{\nu}/\nu) \sim L_{\nu}/\nu^{2}$, the optical depth near $\nu_{\rm pk}$ is thus estimated (at times $t \gtrsim \delta t$) to be
\begin{eqnarray}
&&\tau_{\rm c}(\nu_{\rm pk}) \sim \frac{3}{640 \pi^{2}}\frac{\sigma_T}{m_e}\frac{\nu L_{\nu}|_{\nu_{\rm pk}} \cdot ct\cdot n_{\rm ext}}{\nu_{\rm pk}^{3} r^{2}} \nonumber \\
&\sim& \frac{2\pi^{2}}{405(17-4k)}\frac{m_p}{m_e}f_{\xi}t\nu_{\rm pk}
\sim 5\times 10^{3}f_{\xi,-3}\left(\frac{\nu_{\rm pk}}{{\rm GHz}}\right)t_{-3}
\label{eq:tauC2}
\end{eqnarray}
where in the second line we have used equations (\ref{eq:BM}), (\ref{eq:nuFRB}), and the fact that $\sigma_T \equiv 8\pi e^{4}/(3c^{4}m_e^{2})$.  The scattering optical depth is therefore generically large near the SED peak when the latter is in the range relevant to FRB emission.  

Naively, then, Compton scattering appears to simply increase the effective value of the minimum cut-off frequency from $\nu_{\rm min}$ to $\nu_{\rm pk} \sim 3\nu_{\rm min}$, while at frequencies $\gg \nu_{\rm pk}$ radiation could still escape the upstream.  However, this does not account for the fact that, as the beam is attenuated, {\it the peak of the SED will move to higher frequencies}, thereby increasing the range of frequencies with a positive photon slope that give rise to $\tau_{\rm c} > 0$ and will experience strong scattering.  Although the details of this process are complex and beyond this scope of this paper, we can crudely estimate its effect by adopting the difference equation (\ref{eq:tauC2}) as an estimate of the effective optical depth at all frequencies up to where scattering becomes ineffective (once $\tau_{\rm c} \lesssim 3$).  In other words, we take
\be
\tau_{\rm c}(\nu) \sim \left(\frac{3}{640 \pi^{2}}\frac{\sigma_T}{m_e}\frac{\nu L_{\nu}|_{\nu_{\rm pk}} \cdot ct\cdot n_{\rm ext}}{\nu^{3} r^{2}}\right) \sim \tau_{\rm c}(\nu_{\rm pk})\left(\frac{\nu}{\nu_{\rm pk}}\right)^{-4},
\label{eq:tauC3}
\ee
where in the final equality we have approximated the spectrum as $\nu L_\nu \propto \nu^{-1}$ for $\nu \gtrsim \nu_{\rm pk}$ (Fig.~\ref{fig:SED}).
Equation (\ref{eq:tauC2}) shows that the optical depth decreases below a value $\tau_{\rm c}$ above the frequency
\be
\frac{\nu}{\nu_{\rm pk}} \approx 6.4\left(\frac{\tau_{\rm c}}{3}\right)^{-1/4}f_{\xi,-3}^{1/4}\left(\frac{\nu_{\rm pk}}{{\rm GHz}}\right)^{1/4}t_{-3}^{1/4}
\label{eq:numintau}
\ee

Once the shock propagates to where the optical depth ahead of it is sufficiently low for radio emission to escape ($\tau_{\rm c} \lesssim 3$; see below), the observer is typically observing the SED of the maser emission (Fig.~\ref{fig:SED}) at frequencies $\sim 3-10$ times above the intrinsic (unattenuated) peak $\nu_{\rm pk}$.  Using equation (\ref{eq:numintau}), the frequency peak of the {\it observed} spectrum, $\nu_{\rm max} \equiv \nu(\tau_c = 3)$, thus evolves downward in time.  At times $t \lesssim \delta t$ we have
\begin{eqnarray}
\nu_{\rm max} \propto \nu_{\rm pk}^{5/4}t^{1/4} \propto t^{-\frac{2+7k}{4(8-2k)}} = \left\{
\begin{array}{lr} 
t^{-1/16} & k = 0  \\
t^{-1} & k = 2, \\
\end{array}
\right. 
\label{eq:numax1}
\end{eqnarray}
while at times $t \gtrsim \delta t$ we have
\begin{eqnarray}
\nu_{\rm max} \propto t^{-\frac{2k+7}{4(8-2k)}} = \left\{
\begin{array}{lr} 
t^{-7/32} & k = 0  \\
t^{-11/16} & k = 2 \\
\end{array}
\right. 
\label{eq:numax2}
\end{eqnarray}
where we have used equation (\ref{eq:nuFRBevo}).

Substituting numerical values for $\nu_{\rm pk}$ from equation (\ref{eq:nuFRBevo}),
\begin{eqnarray}
\nu_{\rm max} \approx \left\{
\begin{array}{lr} 
0.75\,\,{\rm GHz}\,\,f_{\xi,-3}^{1/4}E_{43}^{5/32}\dot{M}_{21}^{15/32}\left(\frac{\beta_w}{0.5}\right)^{-45/32}\Delta T_{5}^{-27/32}t_{-3}^{-7/32} & k = 0  \\
1.36\times 10^{5}\,{\rm GHz}\,\,f_{\xi,-3}^{1/4}E_{43}^{-5/16}\dot{M}_{21}^{9/10}\left(\frac{\beta_w}{0.5}\right)^{-15/16}t_{-3}^{-11/16} & k = 2 \\
\end{array}
\right. 
\label{eq:numax}
\end{eqnarray}
This shows that $\sim$ GHz frequency bursts of millisecond duration are a natural prediction of the discrete shell constant density ($k = 0$) scenario.  As we discuss in $\S\ref{sec:observations}$, the temporally decreasing peak frequency is also consistent with the observed downward drifting frequency structure in the sub-pulses of FRB 121102 \citep{Hessels+18} and 180814.J0422+73 \citep{CHIME+19b}.

Before concluding this discussion, we note an additional subtlety in calculating the effective optical depth: $\tau_{\rm c}$ depends on the burst luminosity, which is itself attenuated by scattering.  In a naive picture where we treat the attenuation of the primary beam as an expoential suppression $\tau_{\rm c} \propto \nu L_{\nu} \propto e^{-\tau_{\rm c}}$, the ratio of transmitted to incident (unattenuated) luminosity, $x \equiv L_{\nu}/L_{\nu}(\tau_{\rm c} = 0)$ is determined from the solution to the implicit equation 
\be
\ln{ x} + x\tau_{\rm c} = 0,
\label{eq:DEQ}
\ee
where $\tau_{\rm c}$ is the optical depth (eq.~\ref{eq:tauR}) calculated using the unattenuated luminosity $L_{\nu}(\tau_{\rm c} = 0)$.  An approximate solution, valid for $\tau_{\rm c} \gg 1$, is $x \approx ln(\tau_{\rm c})/\tau_{\rm c}$.  As shown by the solution in Fig.~\ref{fig:suppression}, the reduction in escaping flux is substantially more gradual with increasing $\tau_{\rm c} $ than the usual exponential suppression for a luminosity-independent optical depth.  While equation (\ref{eq:DEQ}) smoothly captures the correct limits $x(\tau_{\rm c} \ll 1) = 1$ and $x(\tau_{\rm c} \gg 1) \sim 1/\tau_{\rm c}$ it is conceptually incorrect in detail, as $\tau_{\rm c}$ represents the rate of photon restribution towards smaller frequencies to the escape rate rather than an exponential suppression of the flux.

In addition to Compton scattering, Raman scattering by the upstream medium can also in principle suppress radio emission from the shock \citep{Lyubarsky08}.  The nominal Raman scattering optical depth can be related to the Compton scattering depth (\ref{eq:tauC3}),
\be
\tau_{\rm r} \approx \left(\frac{\nu}{\nu_{\rm p}}\right)\tau_{\rm c} 
\label{eq:tauR}
\ee
Given that $\nu \gg \nu_{\rm p} \sim \nu_{\rm pk}/(3\Gamma)$ is a necessary condition to observe the synchrotron maser (Fig.~\ref{fig:SED}), the Raman optical depth would appear to greatly exceed the Compton scattering depth in all cases of relevance.  

However, due to Landau damping, Raman scattering is only effective at suppressing the observed pulse if the Debye length in the upstream plasma is sufficiently small that photons are scattered outside of the beam (eqs.~ 19, 27 of \citealt{Lyubarsky08}).  As shown in \citet{Lyubarsky08}, equation (\ref{eq:tauR}) only applies if the temperature of the gas ahead of the shock is sufficiently low
\begin{eqnarray}
&&T_{\rm ext} \ll 320\,{\rm K} \left(\frac{n_{\rm ext}r}{t \nu^{2}} \right)  \nonumber \\
&\approx& 2\times 10^{3}{\rm K}\nu_{\rm GHz}^{-2}E_{43}^{1/4}\dot{M}_{21}^{3/4}\left(\frac{\beta_w}{0.5}\right)^{-9/4}\Delta T_5^{-3/2}t_{-3}^{-3/4},
\label{eq:TextR}
\end{eqnarray}
where in the second equality we have used equation (\ref{eq:rdecdiscrete}), (\ref{eq:ndiscrete2}) for the discrete shell case.  As shown in $\S\ref{sec:afterglow}$, the immediate upstream plasma is heated by Compton scattering from gamma-rays emitted behind the shock to $T \gtrsim 10^{6}$ K (eq.~\ref{eq:TC}).  Because Raman scattering is greatly suppressed at such high temperatures, we are justified in neglecting it relative to Compton scattering.

\subsection{Comparison to FRB Observations}
\label{sec:observations}

\begin{figure}
\includegraphics[width=0.5\textwidth]{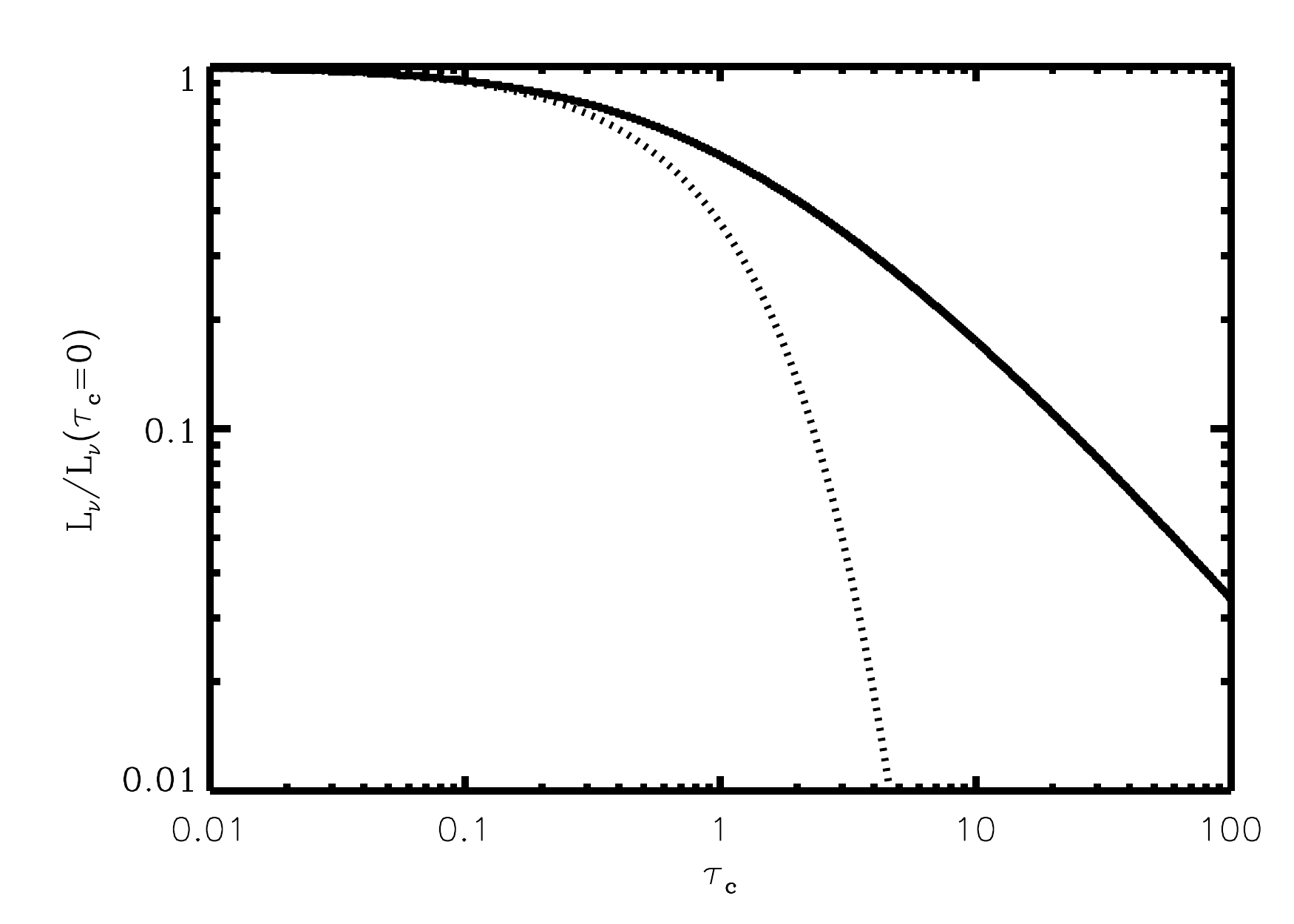}
\caption{A solid line shows the solution to equation (\ref{eq:DEQ}) for the ratio of the transmitted to incident luminosity due to attenuation by induced Compton scattering, as a function of the optical depth $\tau_{\rm c}$ (eq.~\ref{eq:tauC}) calculated using the incident (unattenuated) luminosity, $L_{\nu}(\tau_{\rm c} = 0)$.  Shown for comparison with a dashed line is the naive exponential $e^{-\tau_{\rm c}}$ suppression for the standard case of luminosity-independent $\tau_{\rm c}$.}
\label{fig:suppression}
\end{figure}

The bolometric luminosity of the intrinsic (unattenuated) maser emission (eq.~\ref{eq:Lnu}) is controlled, in both steady-wind and discrete ejecta shell scenarios, by the properties of the ultra-relativistic ejecta ($E$, $\delta t$) and the shock radiative efficiency $f_{\xi}$.  
However, the optical depth of the upstream medium near the peak of the SED $\nu_{\rm pk}$ is generally enormous (eq.~\ref{eq:tauC2}), completely attenuating the signal.  Nevertheless, as the shock moves outwards through the external medium, the value of $\tau_{\rm c}$ at a given frequency drops monotonically.  The timescale and other properties of the observed radio emission are therefore generally set by the epoch at which $\tau_{\rm c}$ first reaches values $\lesssim 3$.  As shown by equation (\ref{eq:numintau}), when this occurs the observer frequency obeys $\nu \gtrsim 10 \nu_{\rm pk}$, while Fig.~\ref{fig:SED} shows that only a fraction $f_{\nu} \lesssim 10^{-2}$ of the bolometric maser power is released at these high frequencies.  Accounting also for the fact that the shock energy is shared over several decades in time, the {\it effective} efficiency of FRB emission in the observing band is therefore typically $\sim f_{\nu}\cdot f_{\xi} \sim 10^{-6}-10^{-5}$ for $f_{\xi} = 10^{-3}$.  Reproducing observed isotropic FRB energies (e.g.~$E_{\rm iso} \sim 10^{38}-10^{40}$ ergs for FRB 121102; \citealt{Law+17}) on burst timescales of milliseconds requires flares of energy $E \gtrsim 10^{43}-10^{46}$ erg. 

Although the above conclusions are largely independent of the nature of the upstream ion medium, the steady-state wind scenario runs into severe problems which disfavor it.   First, equation (\ref{eq:numax}) shows that the peak frequency of millisecond bursts for the fiducial values of $\dot{M}_{21} \sim 0.01-1$ and $\beta_w \sim 0.5$ needed to explain the persistent source and RM of FRB 121102 are typically several orders of magnitude too high compared to observed FRB emission.  Furthermore, the local DM ahead of the shock $\sim 10^{2}-10^{5}$ pc cm$^{-3}$ (eq.~\ref{eq:DMwind}) generally exceeds the total DM of most FRBs and the residual local DM $\lesssim 55-225$ pc cm$^{-3}$ for FRB 121102 \citep{Tendulkar+17} when contributions from the Galaxy, Galactic halo, and interstellar medium are subtracted from the measured DM.  Finally, the predicted rate at which the spectral peak should drift to lower frequencies, $\nu_{\rm max} \propto t^{-\beta}$, with $\beta = 0.5-1$ for $k=2$ (eqs.~\ref{eq:numax1}, \ref{eq:numax2}) is much steeper than the rate $\beta \approx 0.07-0.14$ measured for FRB 121102.  In particular, \citet{Hessels+18} found that sub-bursts of duration $t \sim 0.5-1$ ms drifted downwards in frequency at the rate $d\nu_{\rm max}/dt \approx -0.2\, {\rm  GHz/ms}$ through the band $1.1-1.7$ GHz.

By contrast, the constant density (discrete ejecta shell) model for the upstream medium fits the observations better in several respects.  First, the peak frequency of the emission (eq.~\ref{eq:numax}),
\begin{eqnarray}
\nu_{\rm max} \approx 0.75\,\,{\rm GHz}\,\,f_{\xi,-3}^{1/4}E_{43}^{5/32}\dot{M}_{21}^{15/32}\left(\frac{\beta_w}{0.5}\right)^{-45/32}\Delta T_{5}^{-27/32}t_{-3}^{-7/32}  \nonumber \\
\approx 0.50\,\,{\rm GHz}\,\,f_{\xi,-3}^{1/4}E_{43}^{5/32}t_{-3}^{-7/32}\left(\frac{n_{\rm ext}}{10^{3}{\rm cm^{-3}}}\right)^{15/32}
\label{eq:numax3}
\end{eqnarray}
falls naturally in the range $0.1-10$ GHz of observed FRBs for burst durations of milliseconds and  fiducial parameters for the upstream medium ($\dot{M}_{21} \sim 0.01-1$; $\beta_w = 0.5$) motivated by the persistent source and RM of FRB 121102.  This is true provided that the interval since the last major flare (electron-ion ejection event) obeys $\Delta T \sim 10^{5}$ s, which is indeed similar to the mean interval between the highest fluence bursts from FRB 121102 \citep{Law+17}.
For the same parameters, the local contribution from the upstream shell to the dispersion measure of the burst, DM $\approx 0.1\,{\rm pc\,cm^{-3}}\dot{M}_{21} \beta_w^{-2}\Delta T_{5}^{-1}$ (eq.~\ref{eq:DMdiscrete}), are within the observational constraints on FRB 121102.  A constant density model for the upstream medium also predicts that frequency structure in the SED will drift downwards in time as $\nu_{\rm max} \propto t^{-\beta}$, with $\beta \approx 0.06-0.22$, close to the range measured for FRB 121102 \citep{Hessels+18}.  \citet{Hessels+18} and \citet{CHIME+19b} also find that the downward drift rate is greater at higher frequencies, which is also broadly consistent with the prediction, $d\nu_{\rm pk}/dt \propto \nu_{\rm pk}/t$, of power-law decay.

\begin{figure}
\includegraphics[width=0.5\textwidth]{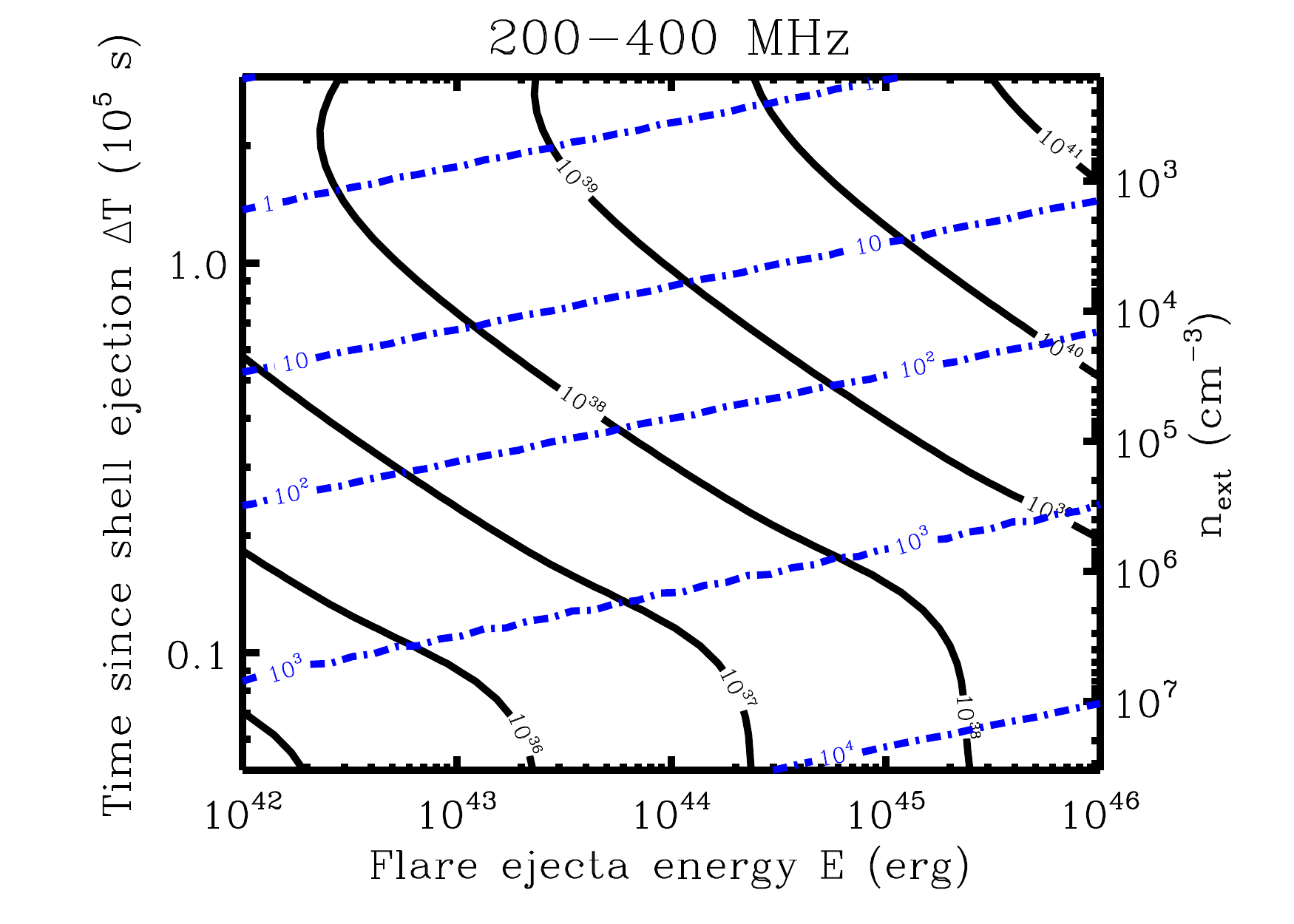}
\includegraphics[width=0.5\textwidth]{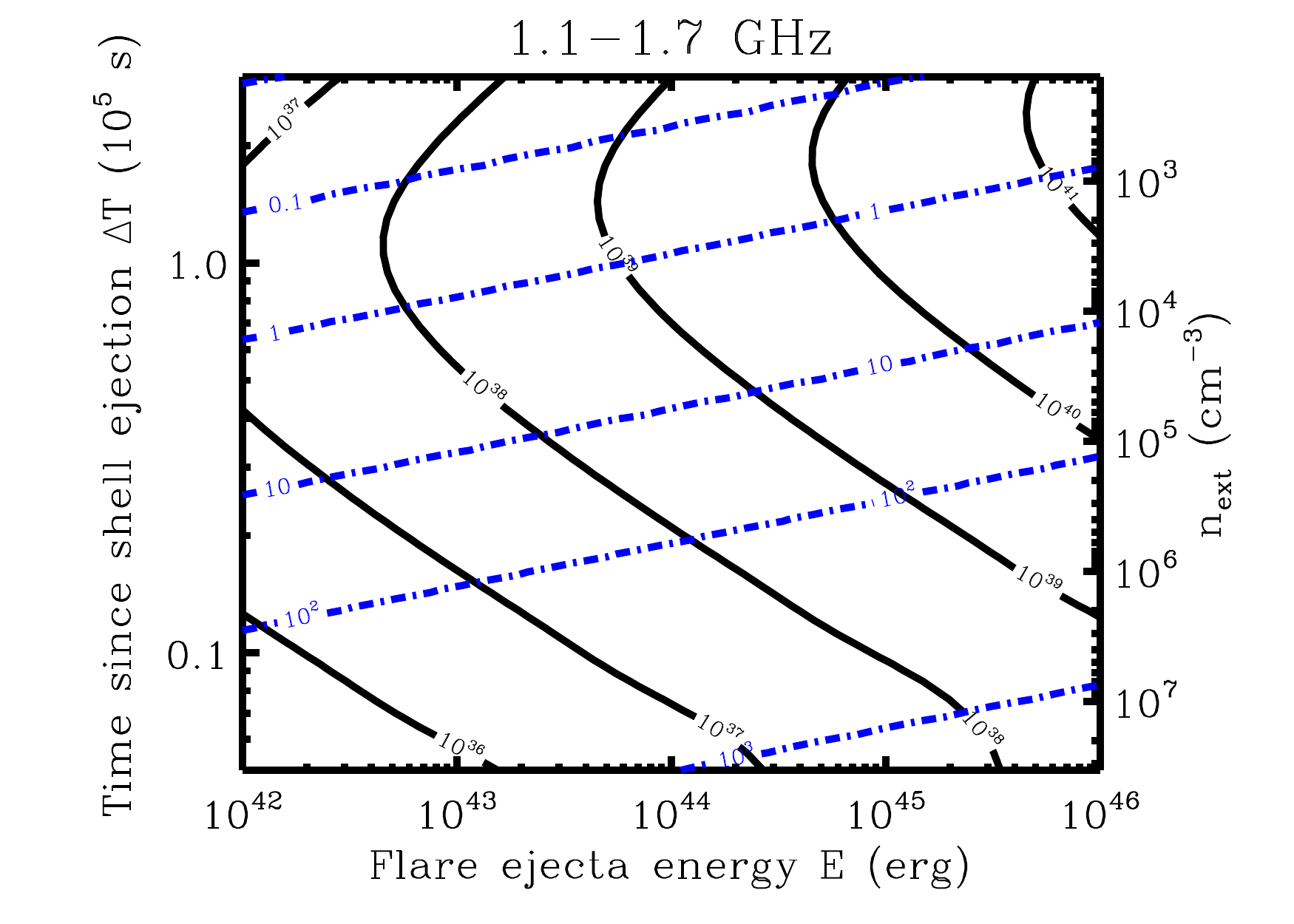}
\includegraphics[width=0.5\textwidth]{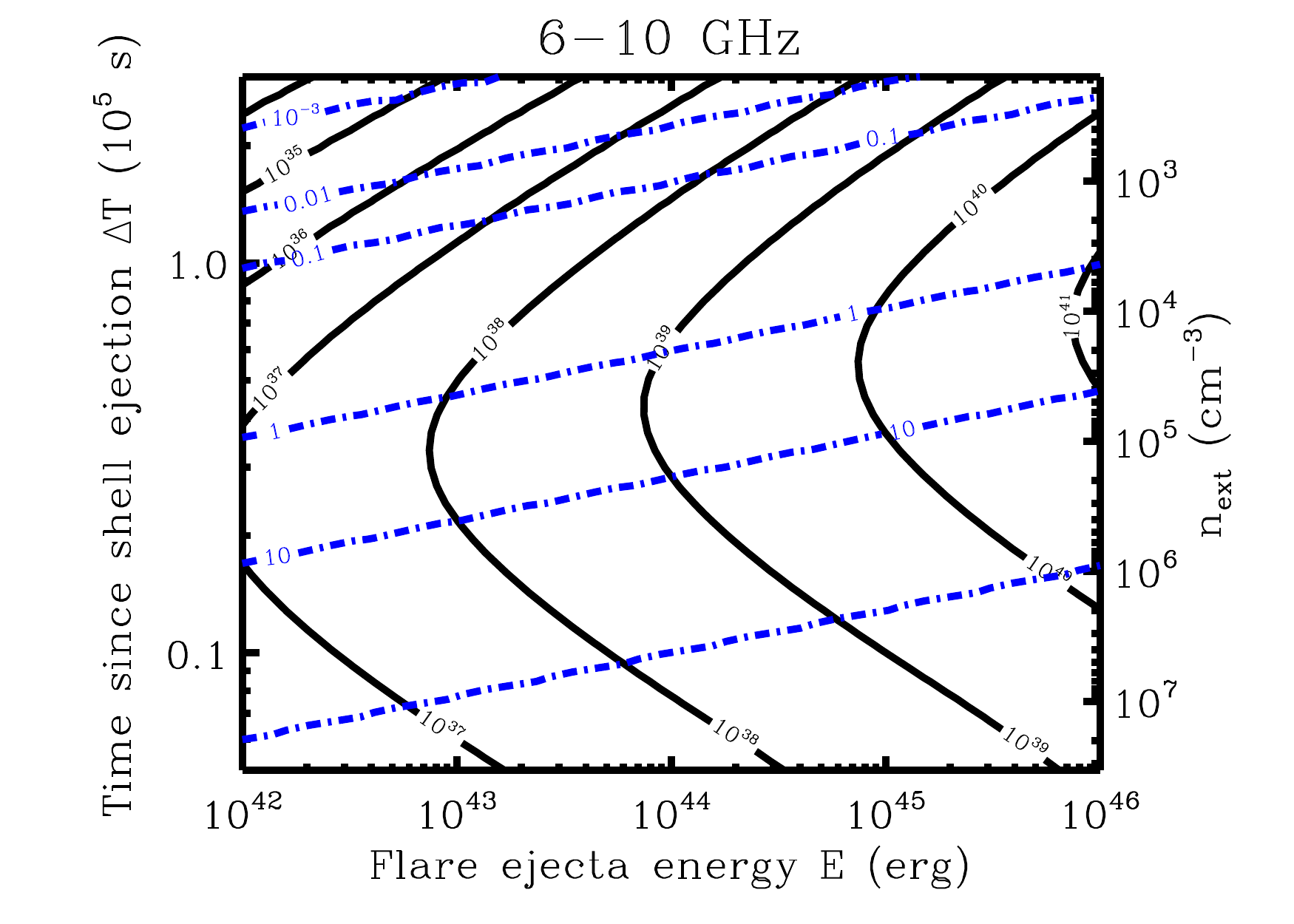}
\caption{Contours of FRB fluence $E_{\rm frb} = \int \int L_{\nu}dt d\nu$ in ergs (solid black lines) and duration $t_{\rm frb} \equiv 3(E_{\rm frb}/L_{\rm max})$ in milliseconds (dashed blue lines) in the space of flare ejecta energy, $E$, and time interval since the last major flare, $\Delta T$.  In calculating the latter we fix the product $\dot{M}\cdot \Delta T = 10^{26}$ g, motivated by the values $\dot{M} \sim 10^{21}$ g s$^{-1}$ \citep{Margalit&Metzger18} and $\Delta T \sim 10^{5}$ s (e.g.~\citealt{Law+17,Nicholl+17}) for FRB 121102;  alternatively, the vertical axis a proxy for the density of the external medium (see right axis).  Different panels show results separately in different observer band-passes: $0.2-0.4$ GHz (top), $1.1-1.7$ GHz (middle) and $6-10$ GHz (bottom).  Increasing flare energy, or the time since the last ion shell ejection $\Delta T$, increases the FRB fluence and decreases the burst duration.  For an otherwise similar flare, the burst duration is shorter, and the fluence greater, at higher observing frequencies.}
\label{fig:contour}
\end{figure}

\begin{figure}
\includegraphics[width=0.5\textwidth]{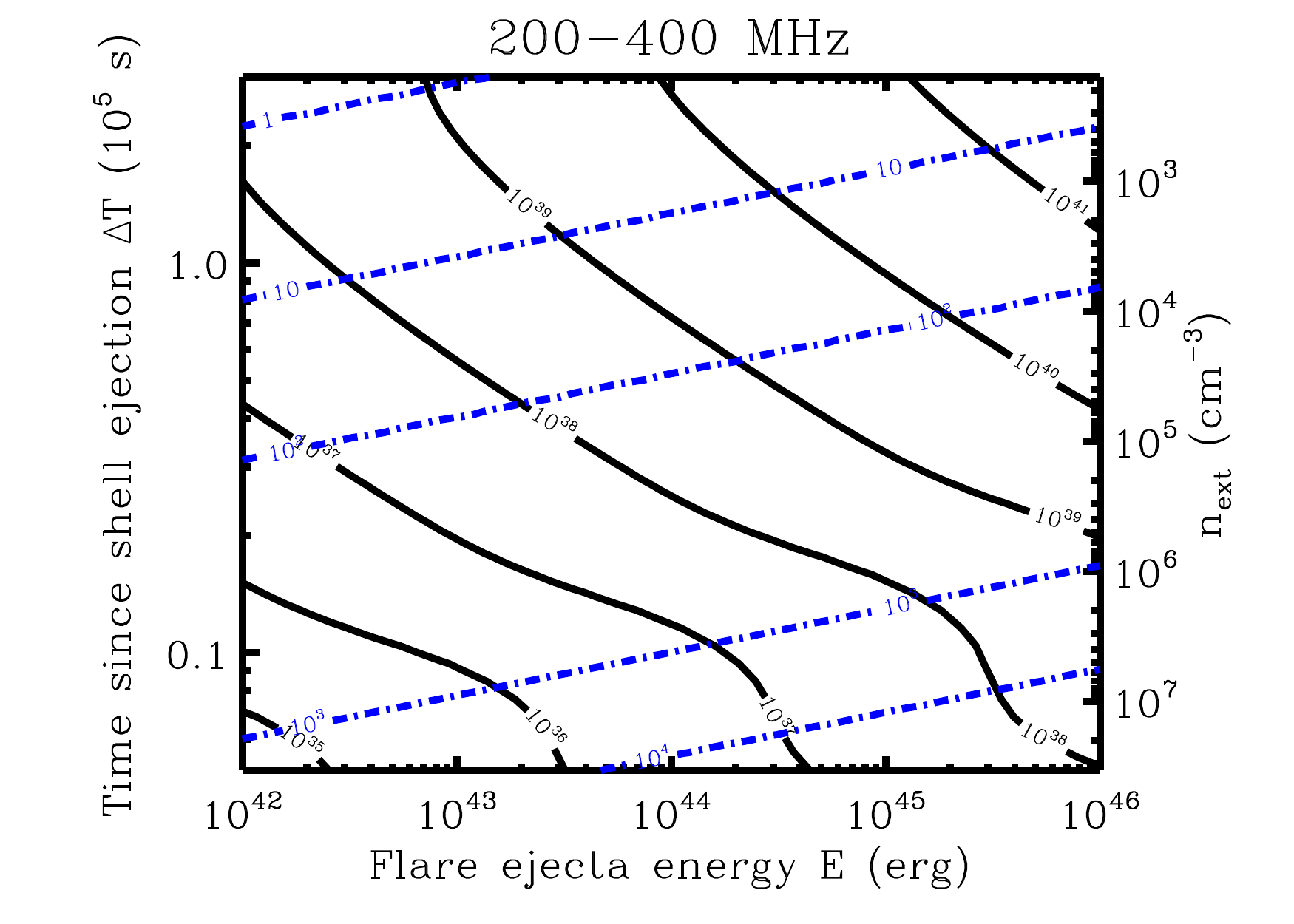}
\includegraphics[width=0.5\textwidth]{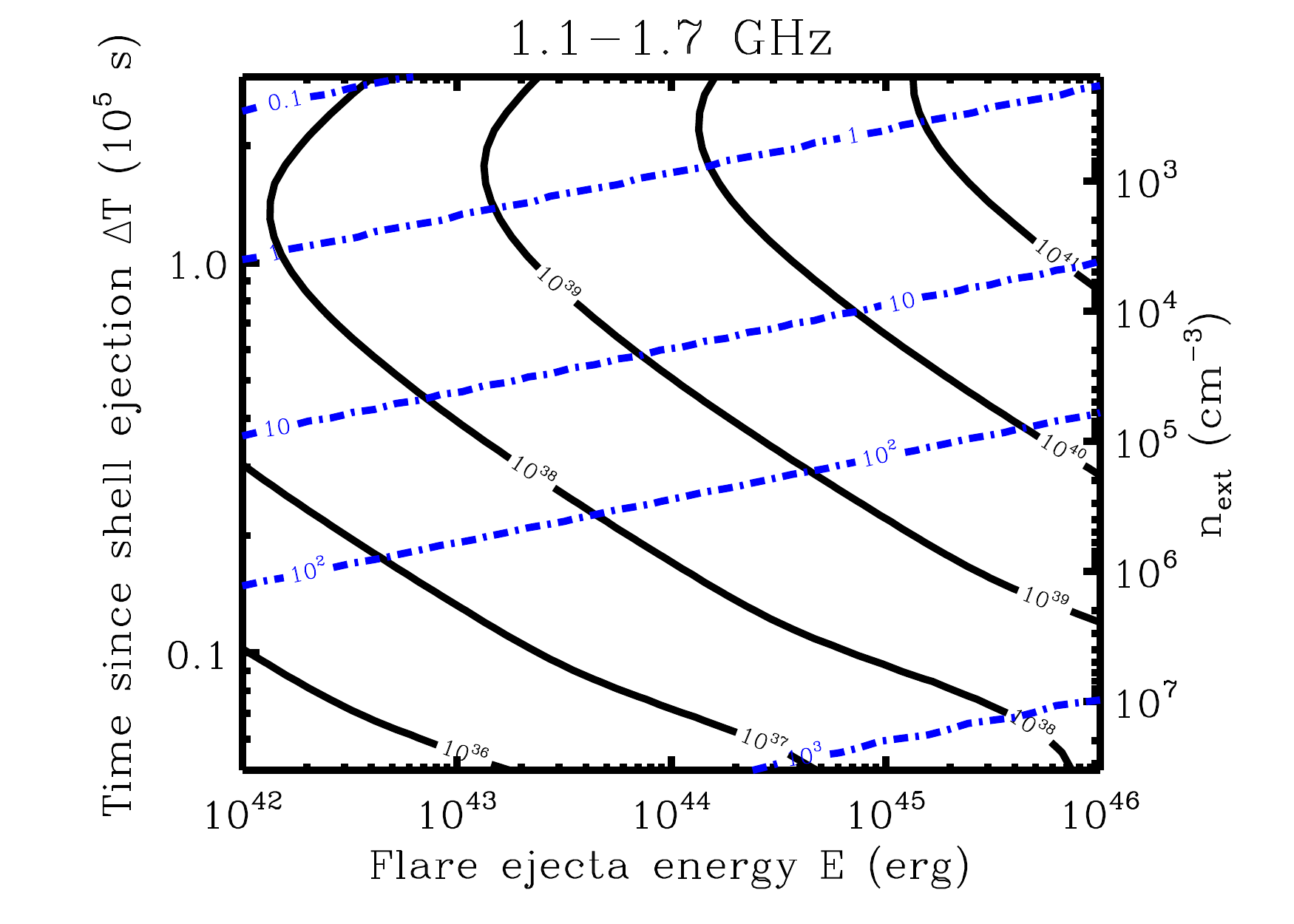}
\includegraphics[width=0.5\textwidth]{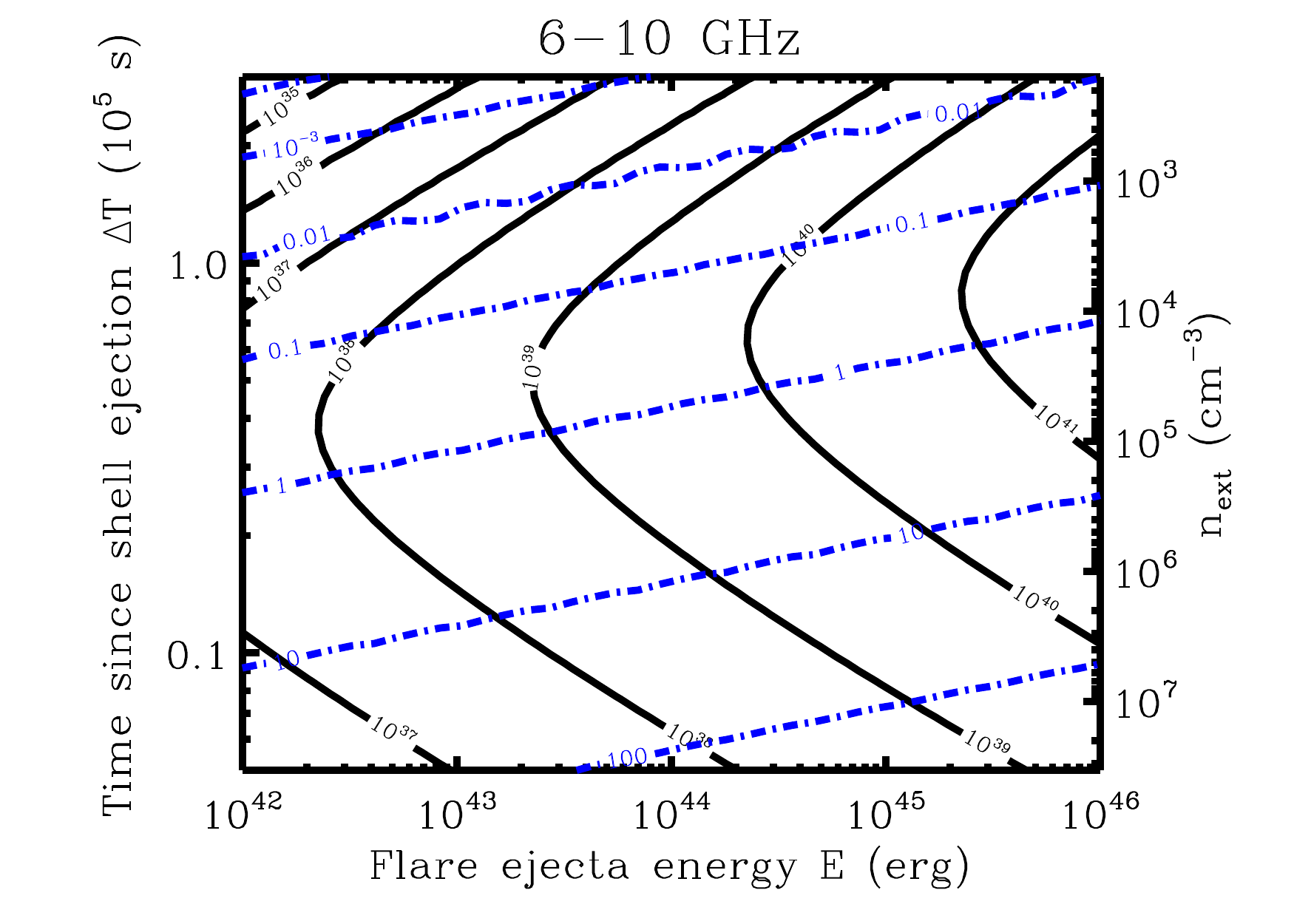}
\caption{Same as Figure \ref{fig:contour}, but with the intrinsic SED of the maser emission calculated for an upstream magnetization $\sigma = 0.3$.}
\label{fig:contour2}
\end{figure}

Figure \ref{fig:lightcurves} shows example light curves of the fluence $F(t) \equiv t \int L_{\nu}(t)d\nu$ in various observing bands (0.2$-$0.4 GHz, 1.1$-$1.7 GHz, 6$-$10 GHz), obtained by combining the predicted time-dependence of the burst luminosity with the predicted SED from \citet{Plotnikov&Sironi19} for different values of the magnetization $\sigma = 0.1-1$ of the upstream medium (Fig.~\ref{fig:SED}).  Here $L_{\nu}$ is calculated accounting for attenuation by induced Compton scattering in the upstream medium using the estimate of the effective optical depth from equation (\ref{eq:tauC3}) combined with Fig.~\ref{fig:suppression}.  Figures~\ref{fig:contour} and \ref{fig:contour2} show contours of the FRB fluence, $E_{\rm frb} \equiv \int \int L_{\nu} dt d\nu $, and duration, $t_{\rm frb}$, as a function of flare energy and $\Delta T$, calculated under the assumption that $\dot{M} \Delta T = constant$, shown separately again for luminosities calculated in the $0.2-0.4$ GHz, $1.1-1.7$ GHz and $6-10$ GHz bands.  Somewhat arbitrarily, we have defined the FRB ``duration" as $t_{\rm frb} \equiv 3E_{\rm frb}/L_{\rm max}$, where $L_{\rm max}$ is the maximum of the bandpass-integrated luminosity $\int L_{\nu}d\nu$.

Consistent with the above estimates, flares of isotropic energies $E \sim 10^{42}-10^{45}$ erg produce bursts with GHz fluences $E_{\rm frb} \sim 10^{36}-10^{41}$ erg and typical durations of $t_{\rm frb} \sim 0.01-10$ ms, compatible with FRB observations, for values of $\Delta T \sim 10^{5}$ s (or, equivalently, external densities $n_{\rm ext} \sim 10^{2}-10^{5}$ cm$^{-3}$).  Notably, the burst duration can be considerably longer than the timescale of the central engine $\delta t = 10^{-4}$ s (e.g. light crossing time of a neutron star) because of the time for the shock to propagate to sufficiently large radii for $\tau_{\rm c} \lesssim 3$, which determines the timescale over which the measured fluence saturates.  In general, the bursts have higher fluence and shorter durations at higher frequencies, at least up until the frequency at which $t_{\rm frb}$ becomes shorter than the intrinsic engine duration $\delta t$.  

Figure \ref{fig:SEDevo} shows snapshots of the SED, which are also calculated by suppressing the intrinsic maser spectrum (Fig.~\ref{fig:SED}) by the frequency-dependent Compton opacity (eq.~\ref{eq:tauC3}).  The combined effects of induced scattering suppression at low frequency, with the fall-off of the intrinsic SED at high frequency, results in the escaping SED peaked at the frequency $\nu_{\rm max}$, which evolves to progressively lower frequency in time (eqs.~\ref{eq:numax1}, \ref{eq:numax2}).\footnote{An additional, narrower peak is observed at lower frequencies, due to radiation that escapes in the relatively low-luminosity gap between $\nu_{\rm min}$ and $\nu_{\rm max}$ (Fig.~\ref{fig:SED}); however, the details of this feature are theoretically uncertain and the suppression may be greater than estimated by equation (\ref{eq:tauC3}) due to the strong dependence of the Compton scattering optical depth on the spectral slope $\tau_{\rm c} \propto (\partial/\partial \nu)(L_{\nu}/\nu)$ (eq.~\ref{eq:tauC}).}  The fine frequency structure shown in this figure at the high-frequency end should also not be over-interpreted; it results in part from the chosen time- and space-sampling of the PIC simulations (which is much shorter than the shock dynamical time over which the observed emission is produced).  Detailed frequency structure will also be washed out by Doppler-broadening effects due to variations in line-of-sight velocity across the emitting shock front.  

The centrally-peaked SED shape (Fig.~\ref{fig:SEDevo}) we predict at high frequencies is broadly similar to those observed in FRB 121102 \citep{Law+17}; in detail, however, for bursts of millisecond duration we typically find a full-width half-max ($\Delta \nu/\nu \sim 1$) larger than those observed, $\Delta \nu/\nu \sim 0.1-0.2$.  This could reflect our highly simplified model for treating the low-frequency suppression from equation (\ref{eq:tauC3}).  The true degree of suppression from Compton scattering is a strong function of the derivative of the photon spectral slope (eq.~\ref{eq:tauC}) and therefore, treated more accurately, would likely create a sharper spectral peak.  Effects associated with plasma propagation could also be playing a role in the observed spectral shape (e.g.~\citealt{Cordes+17}), though this would not be expected to produce a systematic decrease of the frequency structure.

\section{Synchrotron Afterglow}
\label{sec:afterglow}

Behind the forward shock, the particle density and thermal energy density of the hot plasma are given, respectively, by
\be
n_{\rm sh} \simeq 4\Gamma n_{\rm ext}; \,\,\,\, u_{\rm sh} \simeq 4\Gamma ^{2}m_p c^{2}n_{\rm ext}.
\ee
The thermal energy per swept up particle is then given by
\be
\frac{u_{\rm sh}}{n_{\rm sh}} = \Gamma m_p c^{2}
\ee
Non-thermal electrons are not expected to be efficiently accelerated at the quasi-perpendicular magnetized relativistic shocks capable of the synchrotron maser emission \citep{Sironi&Spitkovsky09}.  However, electrons may still be heated, ahead of the shock \citep{lyubarsky_06,Hoshino08} or in the shock layer \citep{Sironi&Spitkovsky10}.  If they are heated to equipartition with the ions, they would achieve a mean thermal Lorentz factor (\citealt{Giannios&Spitkovsky09})
\be
\bar{\gamma} \approx \frac{1}{2}\frac{m_p}{m_e}\Gamma  \gg 1.
\label{eq:gammabar}
\ee
If the magnetization of the upstream medium is $\sigma$, and this compressed field dominates over any shock-generated field, then the magnetic field in the post-shock gas is given in the $\sigma \ll 1$ limit by
\be B = \sqrt{64\pi \sigma \Gamma ^{2} m_p c^{2}n_{\rm ext}}
\ee
The peak frequency of the thermal synchrotron emission (Lorentz-boosted to the observer frame) is thus given by (for the uniform density case, $k = 0$)
\be
h\nu_{\rm syn} = \hbar\frac{eB}{m_e c}\bar{\gamma}^{2}\Gamma  \propto n_{\rm ext}^{1/2}\Gamma ^{4} \propto  \left\{
\begin{array}{lr} 
t^{-1} & t \ll t_{\rm dec}  \\
t^{-3/2} & t \gg t_{\rm dec} \\
\end{array}
\right. ,
\ee
where
\be
h\nu_{\rm syn}(t_{\rm dec}) \approx 57\,{\rm MeV}\,\,\sigma_{-1}^{1/2}E_{43}^{1/2}\delta t_{-3}^{-3/2},
\label{eq:nusyn}
\ee
i.e. the emission is in the gamma-ray band.

The peak maser frequency (eq.~\ref{eq:nuFRB}) and synchrotron peak frequency can be related,
\be
\frac{\nu_{\rm syn}}{\nu_{\rm pk}} = \left(\frac{\sigma}{9}\right)^{1/2}\left(\frac{m_p}{m_e}\right)^{5/2}\Gamma ^{3} \label{eq:nuLnu}
\ee
Thus given $\nu \sim 10 \nu_{\rm pk} \sim$ 1 GHz for the observed FRB, the corresponding peak synchrotron emission at the same epoch is given by $h\nu_{\rm syn} \sim 6\,\,{\rm MeV}\,\,\sigma_{-1}^{1/2}(\Gamma /100)^{3}$ Hz.  A simultaneous measurement of $\nu_{\rm syn}$ and $\nu_{\rm pk}$ would therefore tightly constrain the Lorentz factor of the shock.

In the plasma rest-frame we see that $\nu_{\rm syn}$ (eq.~\ref{eq:nusyn}) is far below the burn-off limit (\citealt{Guilbert+83}),
\be
\frac{(h\nu_{\rm syn})_{\rm max}}{\Gamma}  \sim \frac{9 m_e c^{2}}{\alpha_{\rm F}} \simeq 160\,{\rm MeV},
\ee
above which electrons cool faster than their gyro orbit around the magnetic field, where $\alpha_{\rm F} \simeq 1/137$ is the fine-structure constant.  This implies that incoherent synchrotron radiative losses are safely neglected during the synchrotron maser emission, satisfying this implicit assumption by \citet{Plotnikov&Sironi19}.

Cooling of the electrons is nevertheless important on the dynamical timescale.  The cooling frequency of the electrons behind the shock is given by 
\begin{eqnarray} 
h\nu_{\rm c} = \hbar \frac{eB}{m_e c}\gamma_{c}^{2}\Gamma \approx 9\,\,{\rm keV}\,\,\sigma_{-1}^{-3/2}\left(\frac{\beta_w}{0.5}\right)^{3}\dot{M}_{21}^{-1}t_{-3}^{-1/2}\Delta T_{5}^{2}
\label{eq:nucool}
\end{eqnarray}
where $\gamma_{\rm c} =(6\pi m_e c/\sigma_{\rm T}\Gamma B^{2}t)$  \citep{Sari+98}.  Because $\nu_{\rm syn} \gg \nu_{\rm c}$ the shock-heated electrons will initially be fast-cooling and thus will radiate a large fraction of the shock power.  However, $\nu_{\rm syn}/\nu_{\rm c} \propto t^{-1}$, such that $\nu_{\rm syn} \lesssim \nu_{\rm c}$ after a time
\be
t_{\rm c} \approx 6.4\,{\rm s}\,\,\sigma_{-1}^{2}E_{43}^{1/2}\left(\frac{\beta_w}{0.5}\right)^{-3}\dot{M}_{21}\Delta T_{5}^{-2}
\ee
at which point 
\be
\nu_{\rm syn} = \nu_{\rm c} \approx 0.6\,{\rm keV}\,\sigma_{-1}^{-5/2}E_{43}^{-1/4}(\beta_{w}/0.5)^{9/2}\dot{M}_{21}^{-3/2}\Delta T_{5}^{3}.
\ee
The predicted synchrotron spectrum at $t < t_{\rm c}$ is broad-band, with total luminosity $\nu L_{\nu} \sim L_{\rm sh}/2$ peaking at $\nu \sim h\nu_{\rm syn}$, but extending as a power-law $\nu L_{\nu} \propto \nu^{1/2}$ down to $\nu_{\rm c}$.  Specifically, at times $\delta t \lesssim t \lesssim t_{\rm c}$, 
\be
\nu L_{\nu} \approx  \left\{
\begin{array}{lr} 
L_{\rm pk}\left(\frac{\nu}{\nu_{\rm c}}\right)^{4/3}\left(\frac{\nu_{\rm c}}{\nu_{\rm syn}}\right)^{1/2} \propto t^{1/6} & \nu < \nu_{\rm c} \\
L_{\rm pk}\left(\frac{\nu}{\nu_{\rm syn}}\right)^{1/2} \propto t^{-1/4} & \nu_{\rm c} < \nu <  \nu_{\rm syn} \\
\end{array}
\right. ,
\label{eq:nuLnu_syn}
\ee
where
\be
L_{\rm pk}(t) \approx L_{\rm sh}(t)/2 \approx 10^{45}{\rm erg\,s^{-1}}E_{43}t_{-3}^{-1}
\label{eq:Lpk}
\ee
At times $t \gg t_{\rm c}$ the electrons are no longer fast cooling and the luminosity in all bands will decrease more rapidly with time.

\begin{figure}
\includegraphics[width=0.5\textwidth]{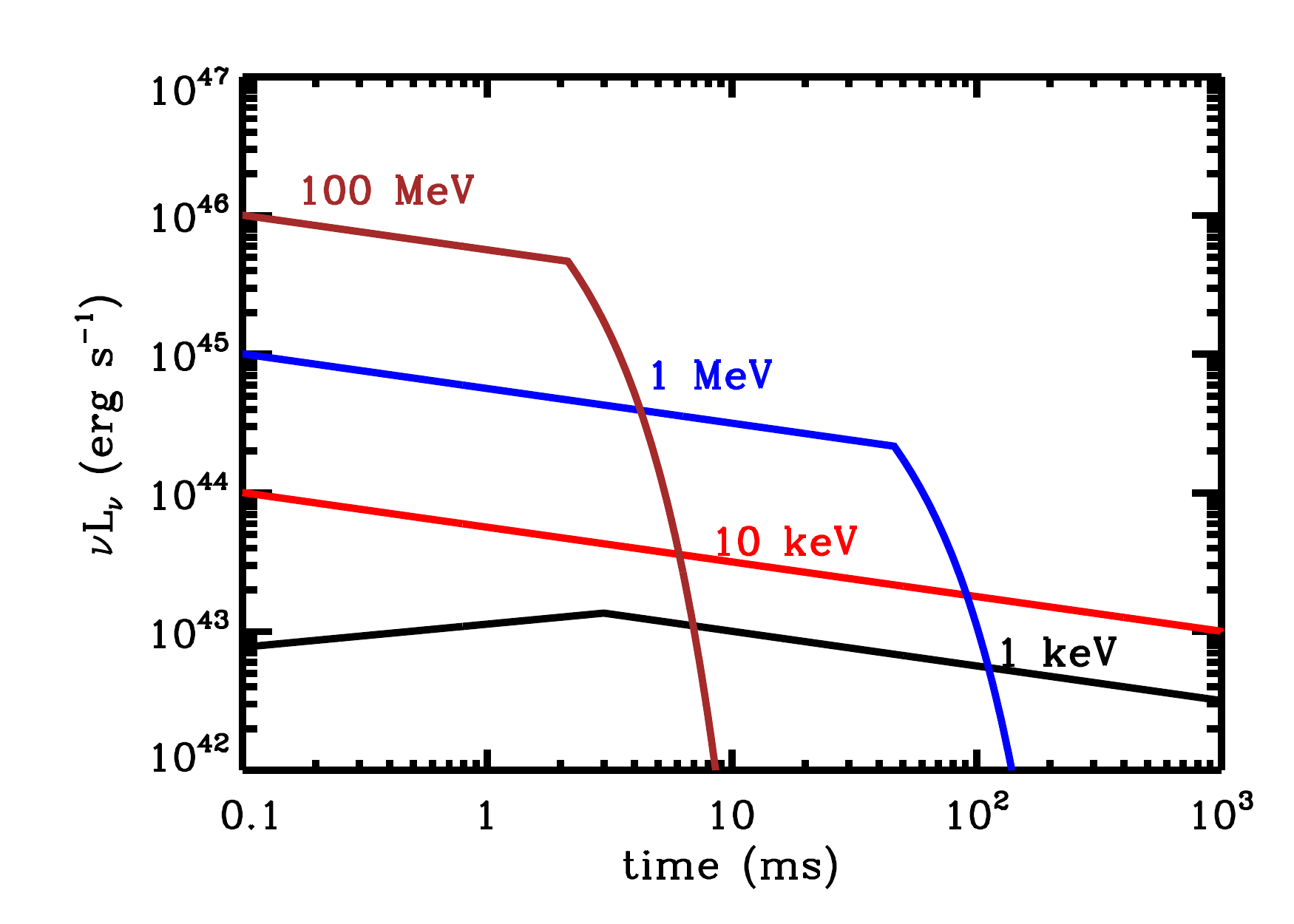}
\caption{Synchrotron afterglow calculated for $\sigma = 0.3$ and otherwise fiducial parameters $\delta t= 10^{-4}$s, $\Delta T = 10^{5}$ s, $\beta_w = 0.5$, $E_{43} = 10$, $\dot{M}_{21} = 1$.}
\label{fig:afterglow}
\end{figure}

Figure \ref{fig:afterglow} shows $\nu L_{\nu}$ light curves for different photon energy ranges from 1 keV to 100 MeV for $\sigma = 0.1$ and other shock parameters matching our $E_{43} = 10$ model from Fig.~\ref{fig:lightcurves}.  Depending on $E_{43} \sim 1-10$ needed to explain typical bursts from FRB 121102, we predict peak luminosities $L_{\gamma} \sim 10^{45}-10^{46}$ erg s$^{-1}$ in the $\sim$ MeV-GeV gamma-ray range on a timescale of $0.1-10$ ms, i.e. comparable to the FRB itself.  The peak luminosity in the 1$-$10 keV X-ray band is typically achieved when $\nu_{\rm c}$ passes down through the observing band at a luminosity $L_{\rm X} \sim 10^{42}-10^{43}\,{\rm erg\,s^{-1}}$ on a somewhat longer timescale of $\sim 0.1-1$ second.  

Unfortunately, these luminosities are below the sensitivity of either gamma-ray or X-ray telescopes at the distance of FRB 121102.  It is also not clear whether the supernova ejecta shell, which surrounds the FRB source in magnetar scenarios, is yet transparent to keV X-rays on the timescale of the radio bursts due to photoelectric absorption \citep{Margalit+18}.  This is consistent with the non-detection of X-ray or gamma-ray counterparts simultaneous with detected bursts from FRB121102 or other FRBs \citep{DeLaunay+16,Scholz+17}. 

Prospects could be better for detecting such ``gamma-ray bursts'' from FRB sources independent of a radio trigger.  Under the assumption that all FRBs originate from repeating sources with a luminosity function similar to FRB 121102, the number density of such FRB sources in the local Universe is $\sim 10^{4}$ Gpc$^{-3}$ \citep{Nicholl+17}.  For the closest source at a distance of $\sim 30$ Mpc the predicted MeV gamma-ray fluence from the most powerful bursts of energy $\sim 10^{44}-10^{45}$ erg would be $\sim 10^{-8}-10^{-9}$ erg cm$^{-2}$, comparable or lower than the weakest GRB (e.g.~\citealt{vonKienlin+14}).  Still, the apparent lack of repeating GRBs in long baseline gamma-ray surveys  (e.g. BATSE) could be used to constrain the high energy tail of the flare distribution, particularly once the repeating fraction among the FRB population is better constrained.

As the visual wavelength band is far below the peak synchrotron or cooling frequencies, optical afterglow emission does not appear to be as promising of a counterpart as that at higher energies.  However, the above analysis assumes the upstream medium is an electron-ion plasma.  If the upstream were of electron/positron composition (e.g. from a rotationally-powered component of the magnetar wind that occurs between magnetic flares), with a number of pairs per ion $(n_{\pm}/n_{\rm ext}) \gg 1$, then $\nu_{\rm pk}$ would be smaller than the above estimate by a factor of $\propto (n_{\pm}/n_{\rm ext})^{-2}$.  Furthermore, the cooling frequency is much lower immediately after a major flare (short $\Delta T$) when the external density is higher.   The best current upper limit on optical radiation simultaneous with a burst from FRB 121102 of $\nu L_{\nu} \lesssim 10^{47}$ erg s$^{-1}$ on timescales $\lesssim 70$ ms \citep{Hardy+17} unfortunately do not constrain this scenario.  Future instruments such as HiPERCAM \citep{Dhillon+16} will be more sensitive by a factor of 100 and thus could place interesting limits.

We emphasize that the escape of high frequency synchrotron emission is not subject to the same constraints as the FRB emission that arise from induced scattering.  Flares from the same source which produce no detectable FRB emission (i.e. in periods after major flares where the density of the external medium is too high) should still in principle produce high frequency afterglow emission, assuming it is not absorbed (e.g. by the supernova ejecta shell) on larger scales.

In addition to their direct signal, gamma-rays from the shock can heat the upstream medium via Compton scattering.  For an incident photon spectrum $F_{\nu} \propto \nu^{-1/2}$ that peaks at frequencies $h\nu \gg m_e c^{2}$, the mean thermal energy of electrons at radius $r$ increases at the rate
\be
\frac{d}{dt}\left(\frac{3}{2}kT_{\rm ext}\right) \sim \frac{\sigma_{\rm T}\nu L_{\nu}|_{h\nu \sim m_e c^{2}}}{4\pi r^{2}}.
\ee
Assuming sufficiently early times that $h\nu_{\rm syn} \gg m_e c^{2} = 0.511$ MeV, such that $\nu L_{\nu}|_{h\nu \sim m_e c^{2}} \sim L_{\rm pk}(m_e c^2/h\nu_{\rm syn})^{1/2}$, then, over a timescale $t$, initially cold electrons ahead of the shock are heated to a temperature,
\begin{eqnarray}
T_{\rm ext} &\sim& \frac{\sigma_{\rm T}}{k}\frac{L_{\rm pk}t}{4\pi r^{2}}\left(\frac{m_e c^{2}}{h\nu_{\rm syn}}\right)^{1/2} \nonumber \\
&\approx& 7\times 10^{6}\,{\rm K}\,\,\,\sigma_{-1}^{-1/4}E_{43}^{1/4}\Delta T_{5}^{-1}\beta_w^{-3/2}\dot{M}_{21}t_{-3}^{1/4},
\label{eq:TC}
\end{eqnarray}
where we have used equations (\ref{eq:rdecdiscrete}), (\ref{eq:nusyn}), (\ref{eq:Lpk}).  This high upstream temperature, while sufficient to suppress Raman scattering of the FRB pulse ($\S\ref{sec:FRB}$; eq.~\ref{eq:TextR}), is not sufficient to suppress the synchrotron maser emission (which requires relatively cold electrons $T \lesssim 0.03 (m_e c^{2}/k) \approx 2\times 10^{8}{\rm K}$ to generate the necessary population inversion; Babul et al., in prep).

UV and X-ray radiation from the shock can also photo-ionize neutral gas ahead of the source, such as the ejecta shell from the supernova explosion, potentially increasing the DM of the bursts.  A powerful flare of energy $\sim 10^{44}$ erg, as determined by the fluence $h\nu \lesssim 1$ keV photons (Fig~\ref{fig:afterglow}), could ionize approximately $N_{\rm flare} \sim 10^{52}$ electrons.  Assume the FRB source is confined within a supernova ejecta shell of mass $M_{\rm ej}$, age $t_{\rm age}$, baryon density $n_{\rm ej} = 3M_{\rm ej}/(4\pi R_{\rm ej}^{3}m_p)$ and radius $R_{\rm ej} = v_{\rm ej}t_{\rm age}$.  The inner layers of the ejecta are swept into a shell of density $4n_{\rm ej}$ by the nebula of radius $R_{\rm n} < R_{\rm ej}$.  The timescale for radiative recombination within the shell is then
\begin{equation}
t_{\rm rec} \approx \left( 4 n_{\rm ej} \alpha \right)^{-1}
\approx 30 \, {\rm yr} \, \alpha_{-11}^{-1}\left(\frac{M_{\rm ej}}{10M_{\odot}}\right)^{-1} v_{\rm ej,9}^3 t_{\rm age,9}^3
\end{equation}
where $\alpha = \alpha_{-11}$ 10$^{-11}$ cm$^{3}$ s$^{-1}$ is the recombination rate.

For the high ionization states of oxygen-rich material at characteristic temperatures $T \sim 10^{4}$ K of photo-ionized gas, we have $\alpha_{-11} \sim 1$, while for O$\textsc{i}$-O$\textsc{ii}$ the recombination rate is significantly lower, $\alpha_{-11} \sim 10^{-2}$.  The recombination time is therefore shorter than the system age $t_{\rm rec} \lesssim t_{\rm age}$ for young sources $t_{\rm age} \lesssim 10-30$ yr such as that estimated for FRB 121102.  In such a case, the number of photo-ionized electrons in steady-state is determined by the number of ionizing photons produced within the recombination time, $N_{\rm p-i} \sim N_{\rm flare} t_{\rm rec} / \Delta T$, where $\Delta T$ is the interval between major flares.  The DM contributed by the photo-ionized layer is therefore given by
\begin{eqnarray}
&&{\rm DM}_{\rm p-i} \approx \frac{N_{\rm p-i}}{4\pi R_{\rm n}^2} \nonumber \\
&\sim& 0.2 \, {\rm pc \, cm}^{-3}
\, \alpha_{-11}^{-1} \left(\frac{M_{\rm ej}}{10M_{\odot}}\right)^{-1} v_{\rm ej,9}\Delta T_{5}^{-1}
t_{{\rm age},9}\left(\frac{R_{\rm n}}{R_{\rm ej}/3}\right)^{-2}.
\label{eq:DMpi}
\end{eqnarray}
For an oxygen-dominated composition, the relevant recombination time is the average over ionization states $\alpha^{-1} = \sum_i \alpha_i^{-1} / 8 \sim 100$ as dominated by O$\textsc{i}$-O$\textsc{ii}$ with the lowest recombination coefficients ($\alpha_{-11} \sim 10^{-2}$).  Although this estimate is at best accurate to an order of magnitude, we find ${\rm DM}_{\rm p-i} \sim 10\, {\rm pc \, cm}^{-3}$ for a source age $t_{\rm age} \sim 10$ yr and otherwise fiducial parameters.  

In cases where the magnetar flare activity is constant in time, the photo-ionized DM is predicted to grow linearly with time.  This could in principle contribute to the $\sim 1-3 \, {\rm pc \, cm}^{-3}$ DM growth in DM measured for FRB 121102 over several years if secular in nature \citep{Hessels+18}.  More detailed photo-ionization calculations are needed to improve these estimates, including contributions to the ionizing flux a rotationally-powered component of the magnetar wind \citep{Margalit+18}.

\section{Discussion and Conclusions}
\label{sec:discussion}

Motivated by recent PIC simulation results \citep{Plotnikov&Sironi19}, we have explored the implications of synchrotron maser emission at magnetized relativistic shocks as a mechanism for fast radio bursts, as first described by \citet{Lyubarsky14} and \citet{Beloborodov17}.  The shocks are generated by the deceleration of ultra-relativistic shell of energy, likely produced by a central compact object, by a dense external environment.  One significant difference from previous work is our assumption that the external medium is a  sub-relativistic electron-ion outflow, instead of an ultra-relativistic wind.  This is motivated by the high injection rate of electrons needed on larger radial scales to explain the observed persistent synchrotron emission and high rotation measure of FRB 121102, assuming both properties arise from the same compact nebula \citep{Margalit&Metzger18}.  

Our main conclusions are summarized as follows:
\begin{itemize}
\item{The shock-powered synchrotron maser as an FRB emission mechanism is consistent with a number of observations, including high intrinsic linear polarization and a spectral energy distribution with complex frequency structure imprinted by high-order harmonics (Fig.~\ref{fig:SED}) and, potentially, by frequency-dependent induced Compton scattering by the upstream medium.  The roughly constant polarization angle of the bursts from FRB 121102 requires an upstream ordered magnetic field with a fixed direction over many bursts.  The latter is naturally expected if the magnetic field is wrapped around the fixed rotation axis of a central compact object.}  
\item{At early times, when the shock is at small radii, the radio pulse is attenuated by induced  scattering in the upstream medium \citep{Lyubarsky08}.  Raman scattering is suppressed by heating of the upstream medium by gamma-rays from the shock (eq.~\ref{eq:TC}), but Compton scattering should be operational.  The observed radio emission only peaks once the shock reaches sufficiently large radii for the induced scattering optical depth $\tau_{\rm c}$ to decrease below values of order unity (Fig.~\ref{fig:suppression}).  For this reason the duration of the FRB for shocks propagating into high density media (small $\Delta T$) can greatly exceed the intrinsic timescale of the central engine (Figs.~\ref{fig:contour}, \ref{fig:contour2}), e.g.~$\delta t \lesssim 10^{-4}$ s if set by the light crossing time of a neutron star magnetosphere.  However, because of the relatively flat fluence curve (Fig.~\ref{fig:lightcurves}), even bursts with total durations of several milliseconds would show significant power on timescales as short as $t \sim \delta t$, consistent with the substructure in FRB 121102 down to 30$\mu$s \citep{Michilli+18}.  

For shocks that propagate into a lower density medium (large $\Delta T$), $\tau_{\rm c} \ll 1$ is achieved at times $\lesssim \delta t$ such that the FRB duration can even be shorter than the engine timescale.  \citet{Scholz+16} notes that the intrinsic widths of the bursts from FRB 121102 of $\sim 3-9$ ms \citep{Spitler+16} are consistently longer than the single-component FRBs detected with the Parkes telescope (all widths $\lesssim 3$ ms), pointing to a key difference between the repeating and non-repeating classes (see also \citealt{Palaniswamy+18}).  We hypothesize that some of the non-repeating population could originate from more powerful flare ejecta propagating into a lower density medium $n_{\rm ext}$ (e.g.~large $\Delta T$; far upper right hand corner of Figs.~\ref{fig:contour}, \ref{fig:contour2}).  The medium could be that surrounding a less active magnetar (long $\Delta T$) or of an entirely different engine.
}
\item{Deceleration of the forward shock, combined with time-dependent attenuation of the radio emission by induced Compton scattering, causes the peak frequency and luminosity of the observed maser emission to decrease as power-laws in time, $\nu_{\rm max} \propto t^{-\beta}$ with $\beta \approx 0.06-0.22$ for $k = 0$ (eqs.~\ref{eq:numax1}, \ref{eq:numax2}).  This provides a natural explanation for the downward evolution of frequency structure seen from bursts in FRB 121102 \citep{Hessels+18} and 180814.J0422+73 \citep{CHIME+19b}.  Matching the observed rate of frequency drift in FRB 121102 requires an external medium with an approximately constant radial density profile ($k \approx 0$), an assumption compatible with other requirements on the burst properties such as their frequency and duration.}
\item{The condition $\tau_{\rm c} \lesssim 3$ for FRB emission is first achieved when the observing frequency is typically $\sim 10$ times higher than the peak frequency of the intrinsic maser emission (eq.~\ref{eq:numintau}).  This guarantees that the observer first sees the high-frequency tail of the SED, such that the effective radiative efficiency for converting flare kinetic energy into coherent radio emission $\sim 10^{-5}$ for a bolometric maser efficiency of $f_{\xi} = 10^{-3}$.  The latter is the range predicted by PIC simulations \citep{Plotnikov&Sironi19} for upstream magnetizations $\sigma \sim 0.1-0.4$ in the range inferred for the nebula on larger scales around FRB 121102 \citep{Vedantham&Ravi18}.

Radio bursts of isotropic energies $E_{\rm frb} \sim 10^{36}-10^{41}$ erg are produced by flares of isotropic energy $E \sim 10^{42}-10^{45}$ erg.   Given the mean repetition time between the strongest FRBs with $E_{\rm frb} \sim 10^{40}$ erg ($E \sim 10^{44}-10^{45}$ erg) of $\Delta T \sim 10^{5}$ s, and an estimated source age $t_{\rm age} \lesssim 30$ yr for FRB 121102, the implied energy budget of the repeater is $\sim (t_{\rm age}/\Delta T)E \sim 10^{49}-10^{50}$ erg, compatible with the magnetic energy reservoir of a magnetar.}
\item{The FRB emission probes the density profile of the upstream ion medium.  For values of $\dot{M} \gtrsim 10^{19}-10^{20}$ g s$^{-1}$ (as needed to explain the persistent synchrotron nebula surrounding FRB 121102), a steady-state $\propto 1/r^{2}$ radial density profile can be ruled out.  More plausibly, the upstream medium is that of a discrete shell ejected following the last major flare (e.g. as supported by the radio afterglow of the 2004 giant flare from SGR 1806-20; e.g.~\citealt{Gelfand+05,Taylor+05}).  Importantly, millisecond GHz bursts compatible with observations are achieved if the interval since the last major flare is $\Delta T \sim 10^{5}$ s (eq.~\ref{eq:numax}; Figs.~\ref{fig:contour}, \ref{fig:contour2}), compatible with the rate of the most powerful bursts from FRB 121102 (\citealt{Law+17}).}
\item{Given the requirement for a sufficiently large time interval $\Delta T$ since the last major flare (low external density) to explain the observed bursts FRB 121102, even a continually flaring FRB source may go through FRB-free ``dark" phases after major flares, consistent with observed long periods of FRB-free activity \citep{Price+18}.  

However, after $\Delta T$ becomes sufficiently large to allow FRB emission to escape, multiple weaker flares in succession could produce clustered bursts by running into the same ejecta shell.  Repetition is in principle possible on timescales shorter than the dynamical timescale at the shock radius $\sim r_{\rm dec}/c \sim 10^{2}-10^{3}$ s, because each flare shocks only a small fraction of the mass of the upstream shell.  }

\item{The time-evolving ion ejecta shell immediately ahead of the shock could contribute stochastic variations in the local DM of the bursts (eq.~\ref{eq:DMdiscrete}) at the level of $\sim 0.01-1$ pc cm$^{-3}$ on timescales of days to months.  This could contribute to observed DM increase of $\sim 1-3$ pc cm$^{-3}$ seen from FRB 121102 over a 4 year baseline \citep{Hessels+18}.  Bursts that occur shortly after major flares (e.g. $\Delta T \lesssim 10^{4}$ s) could produce larger temporary DM increases, but due to attenuation by induced Compton scattering these high DM events might only be detectable at the highest radio frequencies.  It is intriguing to note that the largest DM burst reported by \citep{Hessels+18} was also that detected at the highest radio frequency.  

In addition, X-rays from the shock can photo-ionize the neutral supernova ejecta shell on larger scales, generating a secular flattening or even rise in the DM as the shell becomes progressively more ionized on timescales of the source age (eq.~\ref{eq:DMpi}). }

\item{Our main conclusions are to some extent independent of the identify of the central engine and thus could be compatible with non-magnetar models.  Ultimately, the main requirement to explain FRB121102 within our model is a magnetized environment with a density in the range $n_{\rm ext} \sim 10^{2}-10^{5}$ cm$^{-3}$ (eq.~\ref{eq:numax3} and right hand axis of Figs.~\ref{fig:contour}, \ref{fig:contour2}) over characteristic radial scales $\gtrsim 10^{13}$ cm surrounding the central engine.

For instance, if the engine were an accreting stellar-mass black hole, the requisite ion source for powering the nebula and supplying its high rotation measure could be an outflow from the black hole accretion disk.  However, the high induced scattering depth in steady wind-type scenarios disfavors this model, unless the accretion source were itself intermittent on a timescale $\Delta T \gtrsim 10^{5}$ s.   Likewise, the high required environmental densities could also be consistent with those found in AGN, e.g. if a young magnetar were embedded in the accretion disk of the supermassive black hole.  This scenario would become favored if the luminosity and rotation measure of the persistent source co-located with FRB 121102 are not found to decay on timescales of a few decades, as predicted in the transient nebula picture \citep{Metzger+17}.

Alternatively, if some of the non-repeating FRBs are produced by one-off energy injection events from young compact objects (e.g.~from the delayed collapse of a supramassive neutron star to a black hole; \citealt{Falcke&Rezzolla14}), and the medium surrounding the object is not cleared out by previous flares, then relevant external density would be that from the expanding supernova ejecta,
\be
n_{\rm ext} \approx \frac{3M_{\rm ej}}{4\pi R_{\rm ej}^{3} m_p} \sim 10^{7}{\rm cm^{-3}}\left(\frac{t_{\rm age}}{{\rm yr}}\right)^{-3}\left(\frac{M_{\rm ej}}{M_{\odot}}\right)v_{\rm ej,9}^{-3},  
\ee
where $M_{\rm ej}$, $v_{\rm ej}$ and $R_{\rm ej} \simeq v_{\rm ej}t_{\rm age}$ are the mass, velocity and mean radius of the ejecta at time $t_{\rm age}$ after the explosion.  Depending on the ejecta mass, values $n_{\rm ext} \lesssim 10^{2}-10^{5}$ cm$^{-3}$ are thus achieved on timescales of $t_{\rm age} \sim 10-30$ yr, similar to the timescale over which the ejecta becomes transparent to GHz radio emission (e.g.~\citealt{Margalit+18}).  In the case of magnetars born from the merger of binary neutron stars (e.g.~\citealt{Nicholl+17}), similar densities are achieved substantially earlier (on a timescale $\lesssim 1$ yr) due to the lower ejecta mass and higher ejecta velocities.

On the other hand, an FRB would not be produced by an energy injection event into a normal interstellar medium, as the upstream magnetization would be too low for the shock to produce synchrotron maser emission (the shock would be mediated by the Weibel instability, instead of gyro motion of charged particles about the compressed upstream magnetic field). }

\item{FRBs have now been detected at frequencies as high as 5 GHz \citep{Spitler+18} and 8 GHz \citep{Gajjar+18} and down to 400 MHz \citep{Boyle+18,CHIME+19c}.  Our scenario produces emission across this frequency range.  A given flare's SED peaks first at high frequencies and then $\nu_{\rm max}$ (eqs.~\ref{eq:numax1}, \ref{eq:numax2}) moves to lower frequencies with time  (Fig.~\ref{fig:lightcurves}, bottom panel).  The intrinsic width of the bursts (e.g. after accounting for scattering broadening) should be longer at lower frequencies, as results naturally from the self-similar time evolution of the blast wave deceleration (Figs.~\ref{fig:contour}, \ref{fig:contour2}).  Indeed, there appears to be evidence for longer burst durations at lower frequencies from FRB121102 \citep{Gajjar+18}.

At a given time, the SED is relatively narrowly peaked about $\nu_{\rm max}$ (Fig.~\ref{fig:lightcurves}, bottom panel), as results from the combination of induced scattering by the upstream medium at lower frequencies ($\tau_{\rm c} \propto \nu^{-3}$; eq.~\ref{eq:tauC}) and the drop-off of the intrinsic maser SED at high frequencies (Fig.~\ref{fig:SED}).  While the width of the observed SED we predict $\Delta \nu/\nu \sim 1$ appears to exceed those measured from time-resolve spectra of FRB 121102 (\citealt{Law+17}), our treatment of the frequency-dependence of induced scattering out of the primary beam using an approximate optical depth is the weakest part of our analysis and thus additional work is required to solidify the detailed spectral predictions.  As a general point, the narrow $\Delta \nu/\nu \sim 0.1$ frequency structure of the bursts from FRB 121102 {\it must} arise from the influence of an external medium in our scenario: even an intrinsically narrow SED would be broadened by the differential Doppler shift across the relativistically expanding blast wave.

There is also evidence that the FRB rate at low frequencies $\nu < 700$ MHz is lower than at 1.4 GHz \citep{Karastergiou+15,Rowlinson+16,Burke-Spolaor+16,Caleb+17,Chawla+17,Sokolowski+18}. 
Several mechanisms can suppress low frequency emission \citep{Ravi&Loeb18}, including synchrotron self-absorption \citep{Metzger+17} by the nebula, free-free absorption by the supernova ejecta \citep{Margalit+18}}.  All else being equal, our model predicts that the intrinsic FRB fluence is lower at lower frequencies (Figs.~\ref{fig:contour}, \ref{fig:contour2}), which could also contribute to the lower rate of low-frequency detections.

\item{The same shock responsible for the coherent synchrotron maser emission also produces an (incoherent) synchrotron afterglow, in many ways analogous to a scaled-down version of those which accompany GRB jets (Fig.~\ref{fig:afterglow}).  However, unlike normal GRB afterglows the emission is produced by {\it thermal} electrons heated at the shock rather than a power-law non-thermal distribution (e.g.~\citealt{Giannios&Spitkovsky09}) because magnetized shocks capable of synchrotron maser emission are not favorable sites of non-thermal electron acceleration (e.g.~\citealt{Sironi&Spitkovsky09}).  

For an electron/ion upstream medium, the signal peaks at hard gamma-ray energies on a timescale comparable or shorter to the FRB itself with longer timescale ($\sim$ seconds) emission in the X-ray band.  Unfortunately, for flare energies in the range needed to explain the properties of observed FRBs, this signal is challenging to detect with current gamma-ray and X-ray satellites, even at the estimated distances of the closest repeating FRB source.  Prospects are better in the visual band, but only if the upstream medium of the shock is much higher density (e.g.~in the dark phases right after major flares, which are unlikely to produce detectable FRB emission) or if the upstream medium is loaded with a large number of electron/positron pairs (e.g.~from a rotationally-powered component of the magnetar wind).  }

\item{Our results may have implications for the long-term evolution of FRB emission from newly-born magnetars.  As magnetars age and become less active (e.g.~\citealt{Perna&Pons11,Beloborodov&Li16}), the intervals $\Delta T$ between their major flares could increase.  For otherwise similar flare energies $E$, Figs.~\ref{fig:contour} and \ref{fig:contour2} show that the burst fluence will initially increase and the bursts will get shorter with increasing $\Delta T$ (decreasing external density).  However, once the burst duration comes to match the engine timescale $\delta t$, the effect of a further decrease in density (larger $\Delta T$) is to decrease the observed fluence by pushing the peak of the synchrotron maser $\propto \nu_{\rm p}$ to lower frequencies relative to the observer bandpass.

Thus, flaring magnetars in our model could effectively turn off as FRB sources after a certain age.  Indeed, the giant flares from Galactic magnetars of age $\sim 10^{3}-10^{4}$ yr occur so infrequently (less than once per decade) that the ion shell has time to expand all the way to the nebula termination shock before the next flare \citep{Granot+06}.  An FRB might still be produced in this case as the relativistic flare ejecta interacts with the nebula itself \citep{Lyubarsky14}, but its properties then become sensitive to details such as the radius and density of the nebula.
}

\end{itemize}

\section*{Acknowledgements}
We thank Andrei Beloborodov, Yuri Lyubarsky, Illya Plotnikov, and Indrek Vurm for helpful information and conversations.  We thank the reviewer, Jonathan Katz, for help comments.  BDM and LS acknowledge funding from the National Science Foundation (grant number  80NSSCK1104).  BM is supported by NASA through the NASA Hubble Fellowship grant \#HST-HF2-51412.001-A awarded by the Space Telescope Science Institute, which is operated by the Association of Universities for Research in Astronomy, Inc., for NASA, under contract NAS5-26555.






\end{document}